\documentclass{article}

\usepackage[preprint]{neurips_data_2023}

\usepackage[utf8]{inputenc} 
\usepackage[T1]{fontenc}    
\usepackage{hyperref} 
\usepackage{url}            
\usepackage{booktabs}       
\usepackage{amsfonts}       
\usepackage{nicefrac}       
\usepackage{microtype}      
\usepackage{xcolor}         
\usepackage{colortbl}
\usepackage{tikz}

\usepackage{booktabs}
\usepackage{hyperref}
\usepackage{natbib}
\usepackage{color}
\definecolor{myred}{rgb}{.8,.0,.0}
\definecolor{myblue}{rgb}{.0,.0,.8}
\definecolor{deeppink}{rgb}{1.0, 0.08, 0.58}

\newcommand{\JimSan}{Jim\'{e}nez-S\'{a}nchez}
\newcommand{\AffBCNAIM}{Artificial Intelligence in Medicine Lab (BCN-AIM), Departament de Matemàtiques i Informàtica, \\ Universitat de Barcelona, Barcelona, Spain}

\definecolor{mypurple}{rgb}{.4,.0,.8}

\usepackage{graphicx}
\usepackage{subcaption}
\usepackage{multicol}
\usepackage{multirow}
\usepackage{comment}
\usepackage{amsmath}

\usepackage{listings}
\usepackage{tikz}

\usepackage{array}
\newcolumntype{P}[1]{>{\centering\arraybackslash}p{#1}}

\newcommand{\etal}{\textit{et al.}}

\title{Atlas of MICCAI: 28 Years of Representation, Collaboration, and the Road to Inclusion}

\author{
Amelia \JimSan$^{1}$ \\
\texttt{amelia.jimenez@ub.edu}\\
$^1$\AffBCNAIM \\
\And 
Laura Arbelaez Ossa$^{1}$ \\
$^1$\AffBCNAIM \\
\And 
Joan Perramon-Lluss\`{a}$^{1}$ \\
$^1$\AffBCNAIM \\
\And 
Kaisar Kushibar$^{1}$ \\
$^1$\AffBCNAIM \\
\And 
Vien Ngoc Dang$^{1}$ \\
$^1$\AffBCNAIM \\
\And 
Tanya Akumu$^{1}$ \\
$^1$\AffBCNAIM \\
\And 
Julia A. Schnabel$^{2,3,4,5}$ \\
$^2$School of Computation, Information and Technology, Technical University of Munich, Munich, Germany\\
$^3$Institute of Machine Learning in Biomedical Imaging, Helmholtz Munich, Munich, Germany\\
$^4$School of Biomedical Engineering and Imaging Sciences, King's College London, London, UK\\
$^5$Munich Center for Machine Learning (MCML)\\
\And 
Marius George Linguraru$^{6,7}$ \\
$^6$Sheikh Zayed Institute for Pediatric Surgical Innovation, Children’s National Hospital, Washington DC, USA\\
$^7$ Departments of Radiology and Pediatrics, The George Washington University, Washington, USA\\
\And 
Karim Lekadir$^{1,8}$ \\
$^1$\AffBCNAIM \\
$^8$Instituci\'{o} Catalana de Recerca i Estudis Avançats (ICREA), Barcelona, Spain\\
}

\begin{document}

\maketitle


\clearpage
\begin{abstract}
Scientific knowledge in medical image analysis is shaped by venues such as the Medical Image Computing and Computer-Assisted Intervention (MICCAI) conference, which influence research agendas, professional credibility, and access to opportunity. As AI methods increasingly depend on large-scale data and compute, these dynamics risk further concentrating influence within already well-resourced regions and institutions. Prior meta-research on MICCAI has examined reproducibility, evaluation, and citation practices, but offers limited insight into who participates in and shapes the field. To better understand these dynamics, we conduct a longitudinal bibliometric analysis of the geographic and institutional structure of the MICCAI community, drawing on 9,748 papers published between 1998 and 2025. We find that the USA and China together account for nearly half of all publications, that high-income countries dominate collaboration networks as central and bridging nodes, while lower-income countries remain peripheral, and that participation inequality (Gini index 0.86–0.91 among contributing countries) has declined only marginally over 28 years. We further analyze the evolution of research topics between 2021 and 2025, distinguishing stable core pillars (e.g., segmentation, diagnosis, MRI/CT) from emerging paradigms (e.g., foundation models), and identify a persistent long tail of underexplored application–method and organ–modality intersections. To support transparency and reproducibility, we introduce the Atlas of MICCAI dataset and an interactive dashboard. We contextualize these findings within existing MICCAI equity initiatives, and propose actionable recommendations to reduce participation barriers, build research capacity, and recognize a wider range of scientific contributions.
\end{abstract}

\section{Introduction} \label{sec:intro} 

Scientific knowledge is produced, legitimized, and disseminated through high-impact conferences and journals, which play a central role in determining whose perspectives shape a field. The Medical Image Computing and Computer-Assisted Intervention (MICCAI) conference, as a leading venue in medical image analysis and computer-assisted surgery, is not only a platform for technical exchange but also a powerful mechanism to create epistemic authority, influencing research agendas, and ultimately healthcare policies and clinical practice. Conferences such as MICCAI also provide critical opportunities for visibility, networking and professional advancement. However, access to these opportunities is not equally distributed: an open question remains: who is able to participate, contribute, and be recognized.

Consider two researchers submitting comparably rigorous work: one validates a segmentation method on a large, multi-center benchmark; the other, working with the only imaging equipment available at their hospital, validates a similarly novel method on the 100 cases their institution has collected. To a reviewer calibrated on large-scale benchmarks, the second submission may read as less rigorous, not because the underlying science is weaker, but because the resources available to produce it differ. Fricker describes this kind of mismatch as a \textit{credibility deficit}: a claim is discounted because of who is making it, or the conditions under which it was produced, rather than its actual quality \cite{fricker2007}. A similar effect can apply to the problem being studied rather than the evidence behind it. For example, research on skin conditions in darker skin tones is often treated as peripheral, not because it is less important, but because it falls outside the field's dominant priorities. 

A second, subtler form of exclusion can occur even earlier: researchers who lack access to the tacit knowledge, networks, and submission norms of a field may struggle to participate. This includes not only understanding its technical language but also knowing what makes a competitive submission and which datasets and benchmarks are viewed as credible. Fricker describes this as an \textit{interpretive gap} \cite{fricker2007}, which can prevent researchers from reaching the submission stage at all. As medical imaging and AI increasingly depend on large datasets, specialized equipment, and compute-intensive methods, these barriers may become even more pronounced, particularly for researchers in countries with limited resources and groups that have historically been underrepresented in shaping research directions. Across a community, such structural mismatches between what participation requires and what is unevenly available can quietly determine whose work is seen as credible, and therefore whose questions and priorities come to define the field. As a result, clinical priorities that fall outside the dominant paradigm may remain structurally less visible regardless of their merit \cite{Kuhlman2020Representation}.

Values are not only visible in who gets to participate, but in the technical choices researchers make. Recent work has shown that seemingly technical decisions in medical image analysis, such as the choice between end-to-end and separable learning architectures, are shaped by underlying research values \cite{baxter2025exploring}. Even more specific choices, such as which source dataset to use for transfer learning, are often guided less by systematic evaluation than by heuristics and community norms \cite{lu2025intuitions}. This provides a concrete example of the interpretive gap described above: what counts as a sensible methodological default is itself something researchers learn through exposure to a field's tacit conventions, which is not equally available to everyone. If community norms shape decisions as narrow as dataset selection, they are unlikely to be absent from broader questions of who participates in the field, which problems are prioritized, and whose work is recognized as credible. Understanding these influences therefore requires looking beyond individual technical decisions to the communities and institutions that produce and transmit them.

While prior work has provided important insights into reproducibility \cite{simko2024reproducibility}, evaluation \cite{christodoulou2024confidence}, citation practices \cite{sourget2024citation}, and the role of data within the MICCAI community \cite{balsiger2021miccai}, these studies focus primarily on methodological and reporting aspects of research. Consequently, they provide limited insight into the geographic, institutional, and power structures that shape participation and influence within the community.

To date, no bibliometric study has systematically examined the geographic and institutional composition of the medical image analysis community. Although disparities in representation are widely acknowledged, they remain largely unquantified. In this study, we adopt an epistemological and ethical perspective to provide the first systematic assessment of these patterns. Using MICCAI, the field's largest and most influential conference, as a lens through which to examine participation, collaboration, and agenda-setting, we combine a longitudinal bibliometric analysis of the community and its research landscape with a position-oriented discussion of the implications of these patterns for representation, inclusivity, and epistemic equity. Our main contributions are as follows:
\begin{itemize}
    \item We create and publicly share the \textbf{Atlas of MICCAI (1998-2025)} dataset, containing metadata for 9,748 MICCAI conference papers across 28 years, including publication year, titles, abstracts, author lists, and institutional affiliations, alongside an interactive dashboard for exploration: \url{https://miccai-dashboard.streamlit.app/}.
    \item We characterize \textbf{who participates and how they collaborate}: which countries and institution types are most active, how participation inequality has evolved (Gini, Pielou), and the structure of cross-institutional and cross-regional collaboration networks.
    \item We analyze \textbf{topic evolution} between 2021 and 2025, distinguishing stable core pillars (e.g., segmentation, diagnosis, MRI/CT) from emerging paradigms (e.g., foundation models), and identify a persistent ``long tail'' of underexplored areas and sparse cross-category intersections that highlight actionable research gaps.

    \item We contextualize participation patterns within existing equity initiatives in the MICCAI community and propose \textbf{actionable recommendations} to support more equitable participation across the dimensions of access, capability, and recognition. 
        
\end{itemize}

\begin{figure*}[]
    \centering
    \includegraphics[width=1.\linewidth]{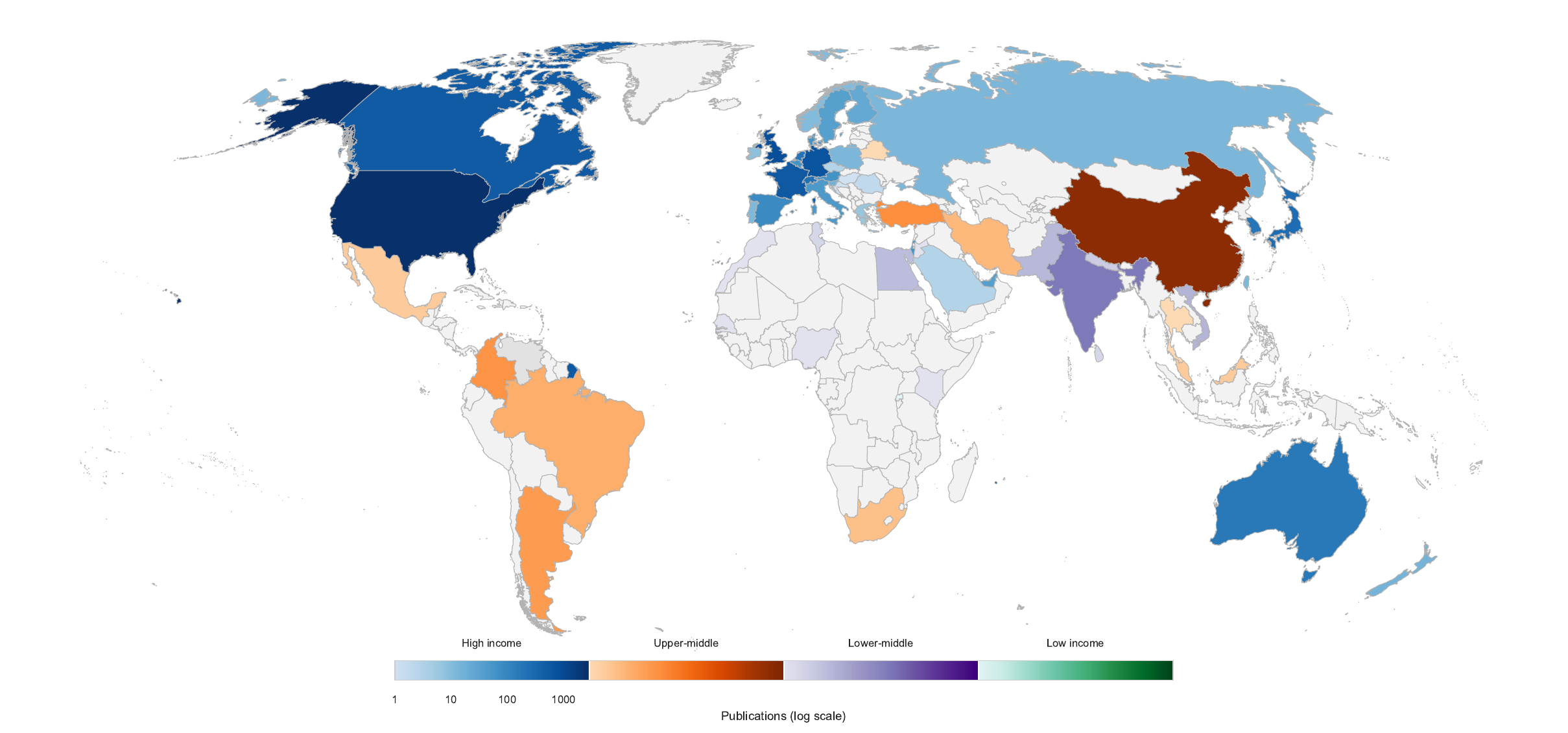}
    \caption{Geographical representation of MICCAI publications by country. The USA and China together account for nearly 50\% of all papers. Overall, 78.99\% of papers originate from \textit{high-income} countries, 19.88\% from \textit{upper-middle-income}, 1.11\% from \textit{lower-middle-income}, and only 0.01\% from \textit{low-income} countries. The hue indicates the country's income category, while color intensity (shown on logarithmic scale) represents the number of first-author papers. }
    \label{fig:worldmap}
\end{figure*}

\section{Background} \label{sec:relatedwork}
\paragraph{Representativeness and Epistemic (In)Justice.} The creation and validation of scientific knowledge is never neutral or context-free: what questions are asked, which agendas are prioritized, and how claims are validated are themselves ethical decisions \cite{harding1991whose}. In medical image analysis, the literature has traditionally been dominated by methodologically and technical contributions, drawing attention to \textit{how} AI is built rather than \textit{what it aims} to solve or \textit{its effects} on clinicians, patients, and societies, with comparatively less attention devoted to questions of clinical utility, implementation, and the societal implications of AI adoption \cite{panayides2025positionjbhi}. For example, many MICCAI benchmarking challenges rely on MRI datasets, even though MRI accounts for a comparatively small amount of real-world imaging data and may present greater barriers to widespread clinical deployment due to its cost and limited availability \cite{ReinkeBenchmarking}. Such benchmarking choices can  incentivize progress on highly standardized technical tasks and leaderboard performance, over what alignment with diverse patients, healthcare systems, and the clinical needs the tools aim to address.

Fricker describes two type of epistemic injustice when knowledge is developed by a limited and structurally skewed set of participants \cite{fricker2007}: \textit{credibility deficits}, where a claim is discounted because of who is making it rather than its quality, and \textit{interpretive gaps}, where researchers lack access to the tacit knowledge, networks, and submission norms that make participation in a field legible in the first place \cite{Kuhlman2020Representation}. Both operate upstream of peer review, shaping who reaches the point of submission at all; authorship analyses in international health research document a related \textit{``stuck-in-the-middle''} pattern, where local partners are recognized as authors but rarely occupy first- or last-author positions \cite{hedt2019stuck}.

Taken together, these accounts suggest that representativeness cannot be reduced to a single headcount. Our empirical analysis examines three dimensions: (i) \textit{presence}, who publishes and participates (Section~\ref{subsec:participationdiversity}); (ii) \textit{collaboration}, who works with whom, through which institutional configurations, and how these patterns change over time (Section~\ref{subsec:collabnetworks} and Section~\ref{subsec:institutionalcollab}); (iii) \textit{agenda-setting}, whose clinical problems and methodological priorities shape the research program (Section~\ref{subsec:trendsInResearchTopics}).

\paragraph{Geographical representation beyond medical imaging research community.} A growing body of bibliometric work shows that scientific fields well beyond medical imaging are dominated by researchers from Western, Educated, Industrialized, Rich, and Democratic (WEIRD) countries. Within Human-Computer Interaction (HCI), Linxen~\etal~\cite{linxen2021weird} found that 84\% of papers came from researchers based in Western countries, with 63\% from the USA alone; Septiandri~\etal~\cite{septiandri2023weird} report a comparably skewed profile for the Fairness, Accountability and Transparency (FAccT) conference. The same imbalance recurs outside HCI and fairness research: Gómez-Cañón~\etal{}~\cite{gomez2025beyond} document a Western-centric authorship paradigm at the International Society for Music Information Retrieval (ISMIR) Conference. In the context of AI ethics, Acu\~{n}a~\etal{}~\cite{acuna2021aiethics} show that, although AI ethics conferences differ from traditional computer science venues (\textit{e.g.}, AAAI, NeurIPS, ICML, ICCV, ACL) in terms of citation dynamics, gender composition, and institutional participation, they still remain characterized by pronounced geographic inequalities.

A related line of work examines not where knowledge is produced, but how the organizations producing it are structured and connected. Vlasceanu~\etal~\cite{vlasceanu2022interdisciplinarity} analyzed 149 organizations with 9,987 affiliated authors publishing at a major AI conference (NeurIPS) and two conferences focused on the societal impacts of AI (FAccT and Artificial Intelligence, Ethics, and Society -- AIES). They found that both academic and industry organizations with influential authors tend to be more interdisciplinary, have a higher fraction of women, and exhibit more hierarchical yet less clustered collaboration structures, even after controlling for organization size.  This institutional lens carries directly into healthcare. Charpignon~\etal~\cite{charpignon2024diversity} examined AI and Critical Care research published on PubMed between 2010 and 2022, assessing author diversity in terms of gender balance, geographic diversity and income level, and the inclusion of researchers from minority-serving institutions located in the United States of America. They find that a greater contribution from low- and middle-income countries tends to co-occur with the use of open critical care datasets. Matos~\etal~\cite{matos2023medical} analyzing co-authorship networks in high-impact medical journals, similarly report that influential authors are disproportionately male and based in high-income countries, with limited growth in collaboration with low- and middle-income regions and evidence of highly connected clusters that concentrate influence. 

Beyond geography and institutional affiliation, a further strand of work argues that representation cannot be understood in isolation from broader power structures and inclusion dynamics. Vethman~\etal{}~\cite{vethman2025intersectionality} adopt an intersectional lens on AI fairness research, calling for interdisciplinary collaboration, reflexivity, and community co-ownership alongside engagement with social and power dynamics. Birhane~\etal{}~\cite{birhane2022values} show, by examining the values encoded in highly cited NeurIPS and ICML papers, how dominant institutions and regions help shape the values of a discipline as a whole. 

While this growing body of work documents geographical imbalances, institutional concentration, and power asymmetries across a wide range of research communities, no prior study has systematically examined these questions within medical imaging. Our work addresses this need by focusing on MICCAI, analyzing the geographic and institutional structure of its research community to characterize patterns of participation and representation at a global scale.

\paragraph{Meta-research at the MICCAI community.} With the advent of deep learning, the MICCAI community has grown substantially, both in size and in the number of scientific contributions. This rapid expansion has brought new challenges: methods have become more difficult to reproduce, and the increasing number of submissions poses questions regarding the selection process and the diversity of research topics. Responding to these concerns, one of the earliest efforts at community-level self-reflection was led by Balsiger~\etal{}~\cite{balsiger2021miccai}, who organized a Hackathon at MICCAI 2020 focused on reproducibility, diversity, and paper selection practices. Their recommendations included immediate and long-term measures that included a mandatory statement of clinical relevance, greater involvement of clinicians, and encouragement of non-standard submissions to increase topic diversity.

Subsequent meta-research has examined the rigor of what actually gets published rather than the selection process itself. Christodoulou~\etal~\cite{christodoulou2024confidence} analyzed all MICCAI segmentation papers published in 2023 and found that most studies report only average performance (\textit{e.g.}, mean Dice score), without confidence intervals or measures of variability. When the authors reconstructed confidence intervals themselves, many ``better'' methods turned out to overlap considerably with others, meaning that headline performance gains are often not statistically distinguishable. A parallel concern is reproducibility itself. Colliot~\etal~\cite{colliot2025reproducibility} argue that reproducibility in medical image computing remains poorly understood and inconsistently assessed, showing through an analysis of MICCAI reviews that reproducibility evaluations are often unreliable and uninformative. They further identify shortcomings in current reviewing practices and advocate for stronger community-wide measures to improve reproducible research. Simko~\etal~\cite{simko2024reproducibility}, examining papers at the related Medical Imaging with Deep Learning (MIDL) conference, report widespread reproducibility issues stemming from missing experimental details such as data splits, preprocessing, and training configurations, with only 22\% of submissions providing a code repository that was deemed repeatable under their evaluation.

A third line of work turns to the datasets underlying this research rather than the methods built on top of them. Heller~\etal~\cite{heller2019role} examined MICCAI papers from 2014 to 2018 and found that over 20\% of studies using publicly available data did not properly cite the corresponding dataset or its associated publication. Sourget~\etal~\cite{sourget2024citation} extend this line of work by analyzing 20 publicly available datasets used in MICCAI and MIDL papers (2013–2023), revealing a strong concentration usage on a limited set of datasets, along with inconsistent citation practices revealing a strong concentration of usage on a limited set of datasets alongside inconsistent citation practices -- a pattern echoed in broader machine learning research \cite{koch2021reduced}, including spikes in publications following high-profile dataset releases such as the 2017 Kaggle lung cancer challenge \cite{varoquaux2022machine, buckeye2017kaggle}. \JimSan~\etal{} further document issues in widely used datasets in computer vision, natural language processing and medical imaging, including missing metadata, unclear licensing, and poor identifier practices, as well as extensive duplication of the ISIC dataset on Kaggle (>600 instances as of their May 2024 study, and >1,500 as of September 2026) that wastes resources and hinders reproducibility. Together, these findings underscore the often secondary status of data contributions relative to methodological innovation \cite{sambasivan2021datawork}.

While this body of work provides important insights into reproducibility, evaluation and citation practices within the MICCAI community, it remains primarily focused on methodological and reporting issues internal to individual papers. Our study addresses a complementary, and, to date, unexplored dimension: the geographic and institutional structure of the MICCAI research community itself, alongside the evolution of its research landscape. Specifically, we investigate diversity across countries, affiliation-type categories, and economic categories, and analyze longitudinal trends in collaboration patterns and research topics, providing a macro-level view of participation and representation that has not been examined in prior work. 

\section{Methodology} \label{sec:methodology}
\subsection{Data collection} \label{subsec:datacollection}
In order to obtain data for our analysis, we implemented a web scraper to collect data from the Conference Proceedings of the largest medical imaging conference: the Medical Image Computing and Computer-Assisted Intervention (MICCAI) from 1998 until 2025. We restricted our analysis to papers from the main conference, without considering workshops. The data collection process consisted of the following steps:
\begin{enumerate}
    \item We gathered all available metadata from the Springer Nature Link website using the Python package Beautiful Soup~\cite{richardson2007beautiful}. We extracted each paper's title and abstract, DOI, list of authors and their affiliations, keywords, and citation metrics. 
    \item We excluded samples corresponding to corrections of other papers, identified by the presence of the words ``Erratum'' or ``Correction to:'' in the title. We also excluded one paper that lacked an abstract. For one paper, we manually input the authors' affiliations, which were missing from the metadata but available in the PDF, and we discarded two papers that did not provide any authors' affiliations. This resulted into a dataset of 9,748 papers. 
    \item We manually curated the affiliation annotations. We generated weak labels using the Llama-4-Scout-17B-16E-Instruct model~\cite{meta2025llama4} to accelerate the process. Specifically, the LLM was used to post-process raw affiliations by removing information about departments and laboratories and retaining only the primary institution name and country, while also proposing a category label for each affiliation based on provided examples. The four possible categories were defined following the Research Organization Registry taxonomy~\cite{researchorganizationresgistry}: \textit{Education} (academic institutions), \textit{Healthcare} (e.g., hospitals, healthcare centers), \textit{Facility} (e.g., research centers, governmental institutions, non-profits, etc.) and \textit{Company}. We then manually verified and refined the automatically generated outputs. We include the prompt used in Appendix \ref{subsec:prompt} for completeness and transparency.
    \item We instructed the LLM to return only one instance per affiliation, thus we manually added additional affiliations (second, third, or fourth) where appropriate. We illustrate this with an example in Table~\ref{tab:affexample}.
    \renewcommand{\arraystretch}{1.2}  

\begin{table}[h]
    \centering
    \begin{tabular}{p{6cm}|p{6cm}}
    Raw affiliation & Mapped affiliations \\
    \hline
    \multirow{4}{6cm}{Univ Rennes, Inria, CNRS, INSERM, IRISA UMR 6074, Empenn ERL U-1228, 35000, Rennes, France}
    & University of Rennes $>$ FRA $>$ education \\
    & INRIA $>$ FRA $>$ facility \\
    & CNRS $>$ FRA $>$ facility \\
    & INSERM $>$ FRA $>$ facility \\
    \hline
    \end{tabular}
    \caption{Affiliation mapping example.} \label{tab:affexample}
\end{table}
    \item We manually corrected anomalies or added missing information. Common hallucinations included incorrect ISO country codes (e.g., MLY instead of MYS for Malaysia) and the use of city codes (e.g., BOS for Boston) in place of country codes (e.g., USA).
    \item We manually standardized institutional affiliations (e.g., converting `Siemens Corporate Research' or 'Siemens Healthineers' to `Siemens', or `Massachusetts Institute of Technology' to `MIT').
    \item We classified each institution's country as \textit{high-income, upper-middle-income, lower-middle-income, low-income}, following United Nations 2026 classification \cite{UNcountryclassification}. For our income-category analysis, we treat Hong Kong, a Special Administrative Region (SAR) of China, as a separate entity, following its classification under the UN's World Economic Situation and Prospects (WESP) report \cite{UNcountryclassification} and the World Bank \cite{worldbank}. Macao SAR (11 papers) lacks a UN WESP classification, so we group it with China. 
    \item We obtained each country-level population and gross domestic product (GDP) from World Bank Open Data repository \cite{worldbank}. 
\end{enumerate}

In total, we obtained the following information for 9,748 papers: authors, title, year, Digital Object Identifier (DOI), abstract, affiliations, institution names, institution types, and institution countries. For analyses involving affiliations, we excluded four papers with missing affiliation information.

\textbf{Research categories.} Since 2021, MICCAI has adopted an open review process, making all accepted papers and their associated categories publicly accessible. We collected papers and category labels from the MICCAI open-access portals for 2021-2025\footnote{\raggedright
\url{https://miccai2021.org/openaccess/paperlinks/categories/index.html} \\  \url{https://conferences.miccai.org/2022/papers/categories/}\\
\url{https://conferences.miccai.org/2023/papers/categories/}\\
\url{https://papers.miccai.org/miccai-2024/categories/}\\
\url{https://papers.miccai.org/miccai-2025/categories/}}. These categories are selected by the corresponding author during submission from a predefined set of options available on the submission portal. The category taxonomy has evolved, categories have been renamed and new categories have emerged over the years. Nevertheless, the four main category groups remain largely consistent across years: \textit{Applications, Machine Learning (Method), Organ/Clinical Application} and \textit{Modality}. To enable longitudinal analyses, we harmonized the (sub-)categories that differed only in naming while retaining the same semantic meaning (e.g., Clinical Application and Organ). The complete list of categories and subcategories per year is provided in Appendix~\ref{subsec:categories_2021_2025}.

To promote transparency and reproducibility, we publicly release the \textbf{Atlas of MICCAI (1998-2025) dataset} in json format via Zenodo\footnote{We plan to openly share our data, but given the risks generative AI poses to unpublished work, we will wait until our manuscript has completed peer review. Please reach out if you are interested in early access.} and make the corresponding analysis available on Github \url{https://github.com/ameliajimenez/atlas-of-miccai} under a CC-BY-NC license. Accompanying the dataset, we provide a Data Card~\cite{Gebru2021datasheets} in the Appendix (Section~\ref{subsec:datacards}) to document its characteristics and intended use. We also provide a dashboard that enables broader exploration of the data by the research community and the public, available at \url{https://miccai-dashboard.streamlit.app/}. 

\subsection{Data analysis}

We conducted the following analysis to investigate the geographical distribution of authorship and institutional affiliations, collaboration networks, and evolution of research topics. 

\textbf{Participation diversity across countries.} To evaluate how research output is distributed across countries, we computed the annual number of publications for each country, along with values normalized by population and gross domestic product (GDP). We then quantified distributional inequality and evenness using the Gini and Pielou indices, respectively. \cite{dixon1987bootstrapping,gini1921measurement} captures inequality in research contributions across countries, whereas the Pielou index \cite{pielou1966measurement} measures how evenly research activity is distributed by normalizing Shannon entropy to capture diversity. Using these indices, we derived linear trends to evaluate whether the distribution of research output has become more balanced over time. We also analyzed year-to-year fluctuations to identify periods marked by pronounced shifts in diversity and concentration.

\textbf{Global collaboration networks.} We constructed bipartite graphs linking authors to their affiliated countries to explore collaboration patterns. Using the Louvain community detection algorithm \cite{blondel2008fast}, we identified clusters of interconnected researchers and institutions. This enables us to evaluate the level of collaboration across regions and examine whether specific research clusters involve affiliations from different economic categories, offering insight into how inclusive these research networks are. In addition, we built country-country collaboration networks to quantify the number of international partners per country and classified publications as either single-country (SC) or multi-country (MC) collaborations. To characterize how countries participate in the global collaboration network, we defined four collaboration profiles: \textit{domestic-oriented} (>60\% SC publications,), \textit{international hubs} (>60\% MC publications with domestic capacity), \textit{balanced} countries (mixed output), and \textit{exclusively collaborative} countries (only MC publications). This typology captures differences in both collaboration intensity and structural position within the global research network.

\textbf{Mapping affiliation types and temporal trends.} We analyzed temporal trends in institutional affiliations by classifying organizations into \textit{Education}, \textit{Healthcare}, \textit{Facility}, and \textit{Company} according to the Research Organization Registry taxonomy~\cite{researchorganizationresgistry}. For each year, we computed the number and percentage of affiliation types per paper. We further applied linear regression to assess trends and identify categories and collaboration groups that are increasing or decreasing over time.

\paragraph{\textbf{Research topics.}} Using the harmonized (sub-)category labels, we computed for each year the number of papers per subcategory, the average number of labels per paper, and the axis coverage, \textit{i.e.} the percentage of papers carrying at least one label in a given main category. Because a paper can have multiple labels within the same main category, the subcategory counts are not mutually exclusive and are reported as absolute counts. We also report label density and coverage to assess annotation completeness across years. In addition, we excluded the 2025 \textit{Deep Learning} label from cross-technique comparisons because it was assigned to 737 papers ($\approx72\%$ of that year's publications), making it too broad to distinguish between techniques. To characterize the long tail, we extracted the five smallest subcategories per main category and year, and separately ranked by mean papers per year those subcategories present in every year from 2021 to 2024, so as to distinguish persistently underexplored topics from artifacts of a single taxonomy. We excluded 2025 from this analysis, since only 30\% of the 2024 labels reappear unchanged the following year, against 65--89\% between previous years. Finally, we quantified cross-category co-occurrence for 2025, the only year with near-universal multi-dimensional tagging, by counting for each pair of subcategories the number of papers carrying both labels.

\subsection{Data visualization}
We provide a dashboard for researchers to explore affiliation types over time, analyze bi-partite networks (author-author and country-country) providing insights into collaboration patterns, and explore paper contributions, affiliation types and collaborators at the country level. The website is available at: \url{https://miccai-dashboard.streamlit.app/}. 

\section{Findings} \label{sec:findings} 
\begin{figure}[t]
    \centering
    \begin{minipage}{0.45\textwidth}
        \centering
        \includegraphics[width=\linewidth]{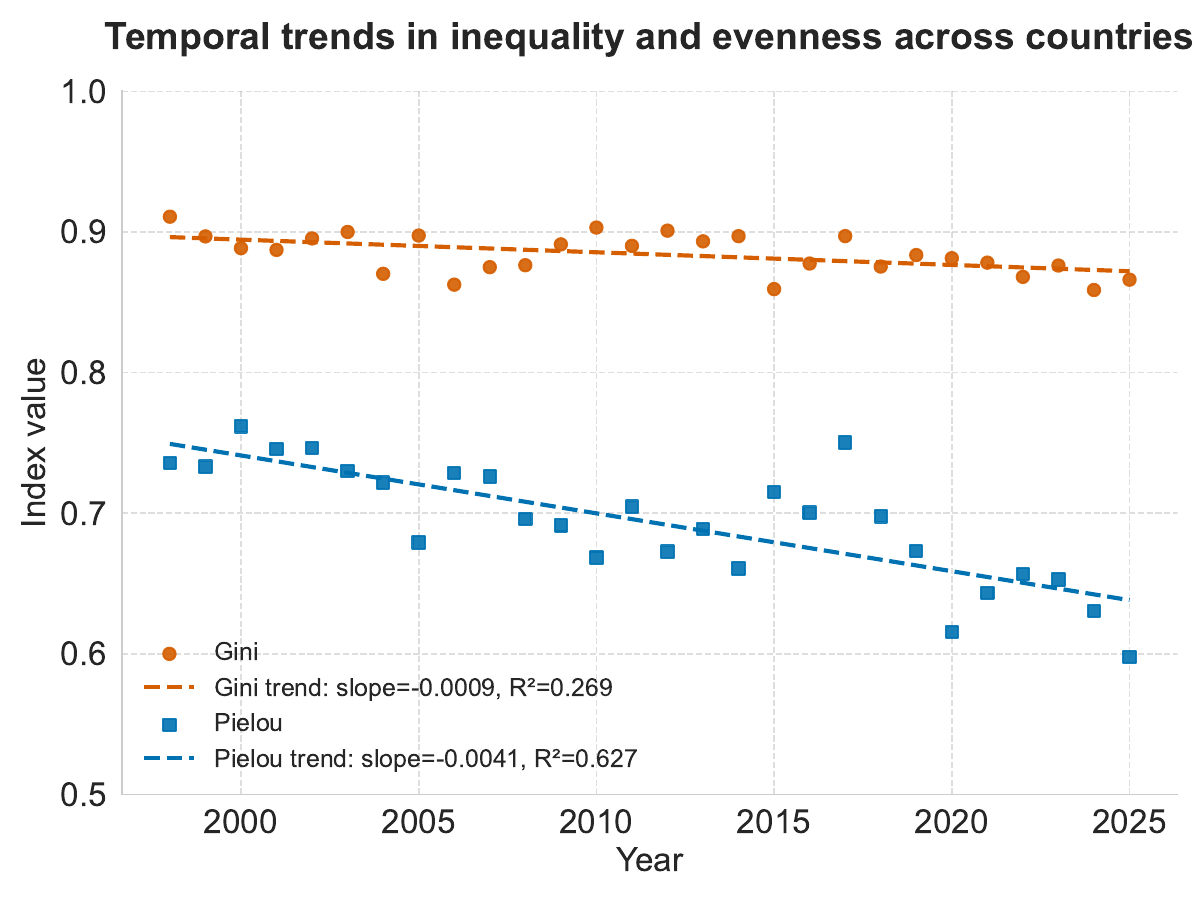}
        \caption{Country distribution over time. The Gini coefficient indicates a very high but slightly decreasing inequality in global participation, while Pielou's evenness shows a stronger decline, suggesting that contributions are becoming increasingly concentrated among a smaller subset of countries.}
        \label{fig:ginipielou}
    \end{minipage}
    \hfill
    \begin{minipage}{0.45\textwidth}
        \centering
        \includegraphics[width=\linewidth]{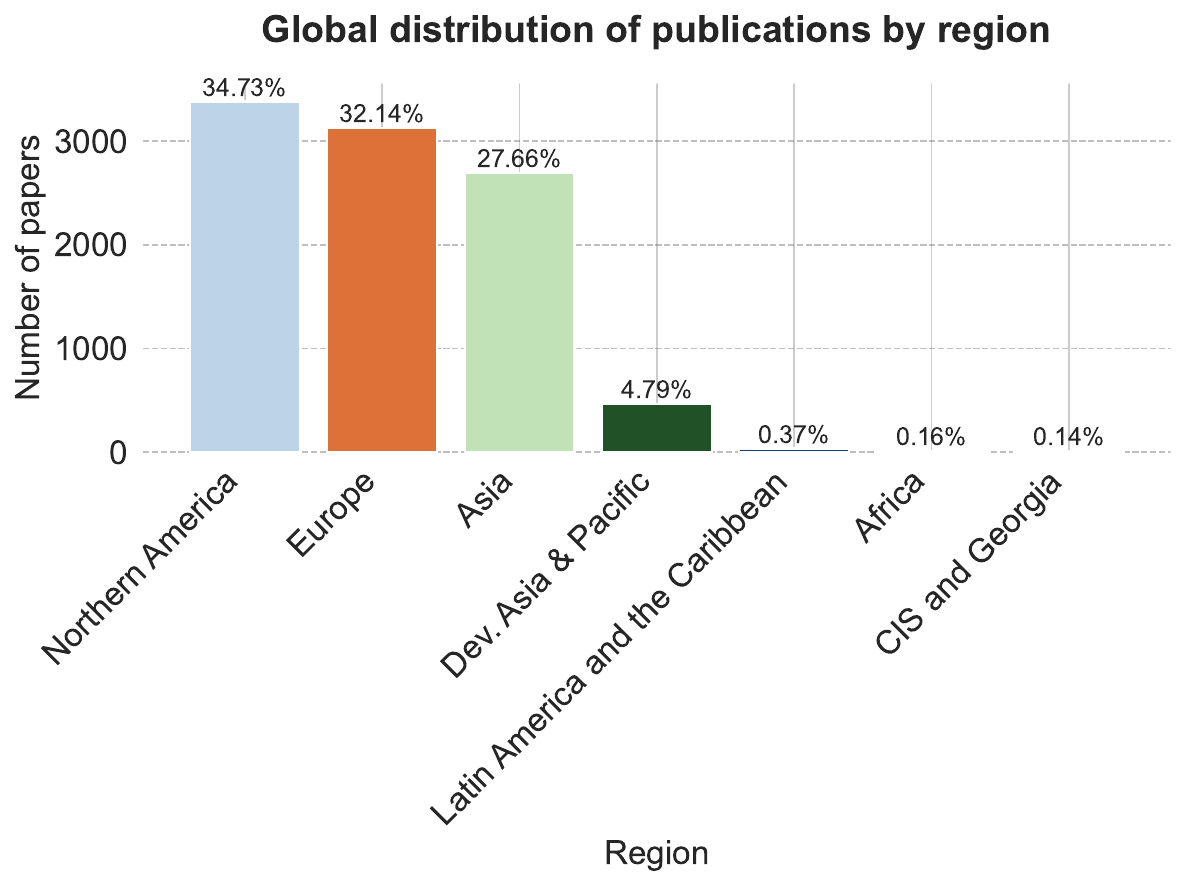}
        \caption{Regional distribution of papers, showing global percentage shares over the years, with notably low representation in Latin America (0.37\%), Africa (0.16\%), and Commonwealth of Independent States and Georgia (0.14\%).}
        \label{fig:regionsbar}
    \end{minipage}
\end{figure}

\begin{figure*}[t]
    \centering
    \includegraphics[width=1.\linewidth]{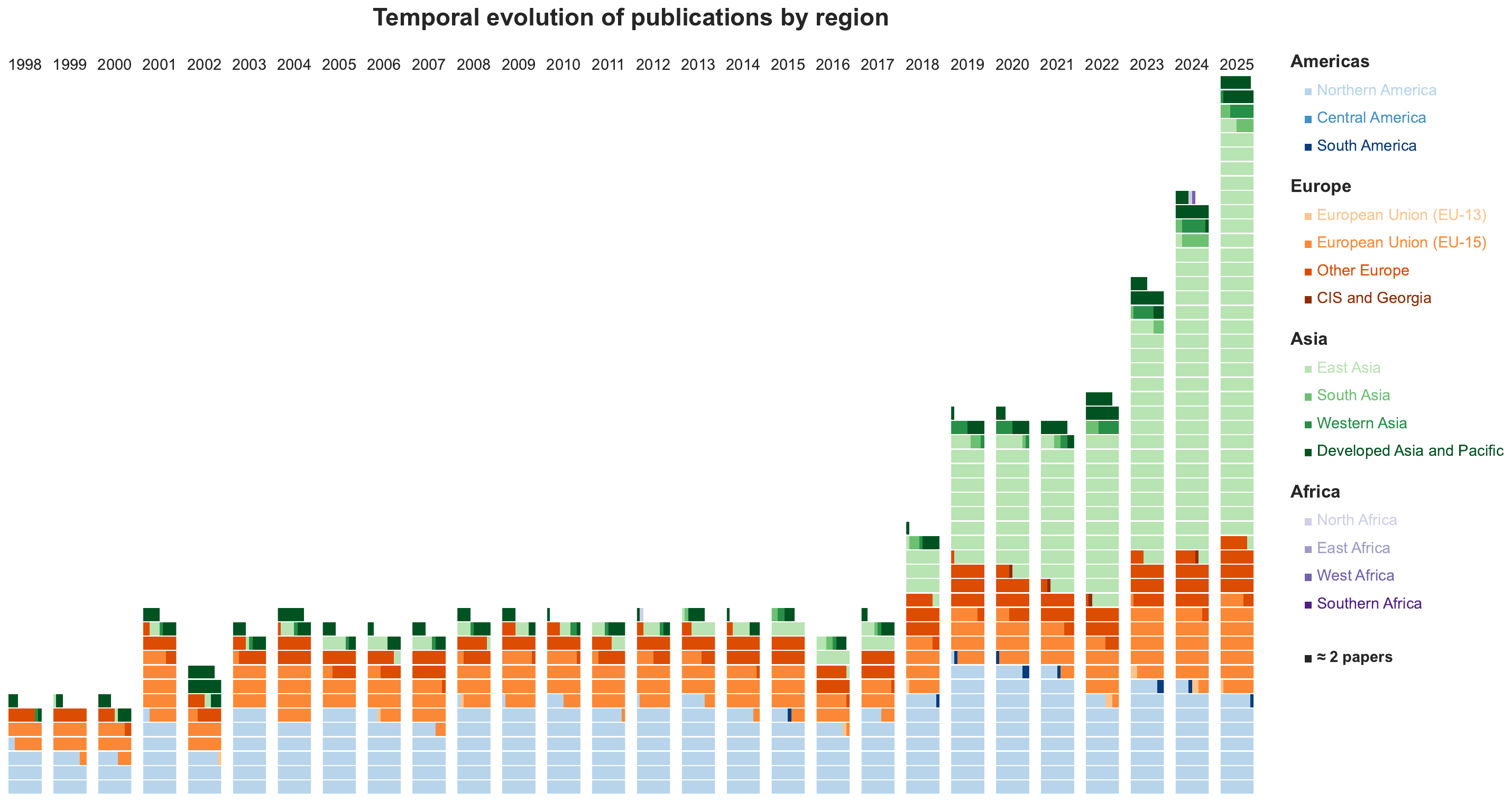}
    \caption{Regional distribution of papers across the years.}
    \label{fig:georep}
\end{figure*}
\begin{figure}[t]
    \centering
        \centering
        \includegraphics[width=\linewidth]{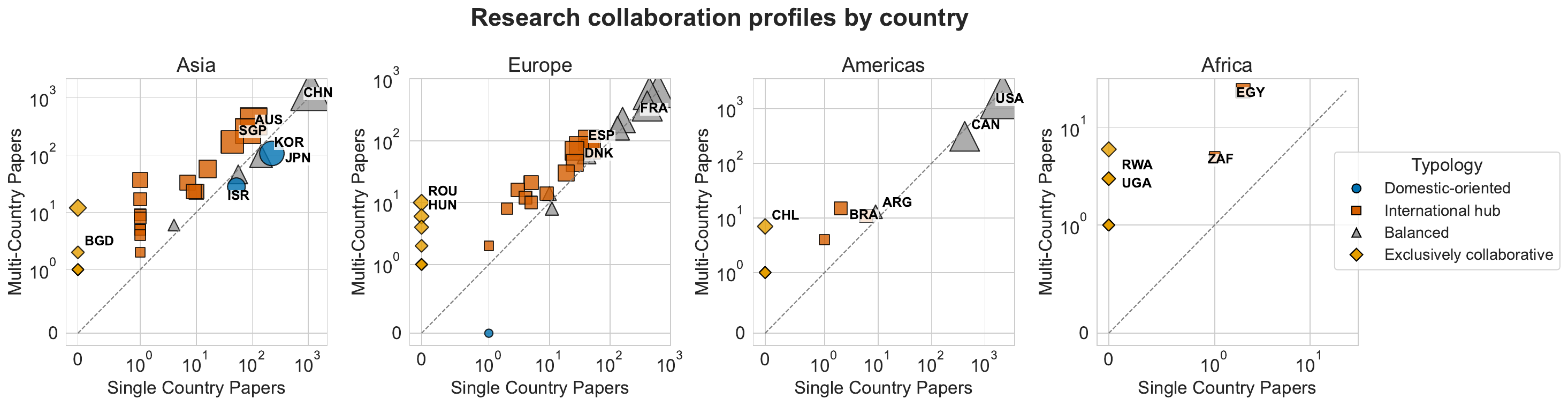}
        \caption{Country collaboration comparison (Single-Country Paper (SCP) \textit{vs} Multi-Country Paper (MCP) for all countries and regional subsets. The panels show the global distribution (left) and regional subsets (Asia, Europe, Americas, and Africa). Countries are classified as \textit{domestic-oriented}, \textit{international hubs}, \textit{balanced}, and \textit{exclusively collaborative}. Marker size indicates total publications. Note the logarithmic scale used to account for large differences in contributions.}
        \label{fig:typologyscpvsmcp}
\end{figure}

\subsection{Participation diversity}
\label{subsec:participationdiversity}
\paragraph{Regional and economic disparities in participation.} Fig.~\ref{fig:worldmap} provides a global overview of country contributions to MICCAI, highlighting both dominant contributors and regions with limited representation. A key finding is the strong dominance of the USA and China, which together account for nearly 50\% of all MICCAI papers. While the USA (2,837 papers) has maintained a relatively stable contribution over time, China (1,877 papers) has experienced a rapid growth in the last years (see Fig.~\ref{fig:georep}). In absolute terms, these countries substantially surpass all other contributors. Looking at contributions normalized by population (Fig.~\ref{fig:worldmap-population}), both the USA (10th of 61) and China (26th of 61) rank well outside the top contributors, which are smaller high-income economies such as Hong Kong SAR, Switzerland, and Singapore, suggesting that population-normalized measures provide a different perspective on participation patterns than absolute publication counts alone.

At the continental level, we observe substantial disparities in participation. Africa is the least-represented continent (16 papers), followed by Central and South America (2 and 34 papers, respectively). Similarly, institutions in Western Asia (128 papers) and Southeast Asia (91 papers) show limited representation, with only a few exceptions. Overall, contributions are distributed as follows: 34.7\% from Northern America, 32.5\% from Asia, 32.1\% from Europe, and only 0.7\% from Africa and Latin America combined (see Fig.~\ref{fig:regionsbar}). These regional disparities are further reflected when grouping contributions by country income level. We find that 78.99\% of papers originate from high-income countries, 19.88\% from upper-middle-income countries, while only 1.11\% and 0.01\% come from lower-middle-income and low-income countries, respectively. 

To contextualize these contributions against economic weight, we compare each income group's share of MICCAI papers to its proportion of global GDP. High-income countries produce 79.03\% of papers while holding 68.16\% of global GDP, a representation ratio of 1.16, slightly higher than their economic share. Upper-middle-income countries are close to proportional (19.84\% of papers vs.\ 26.06\% of GDP, ratio 0.76), while lower-middle-income countries are markedly under-represented: they hold 5.68\% of global GDP but produce only 1.11\% of papers, a representation ratio of 0.20, roughly one-fifth of what their economic weight alone would predict. This disparity suggests that additional structural constraints, including disparities in research funding, scientific infrastructure, and conference accessibility, continue to limit participation from lower-income countries beyond what economic capacity alone would explain.

\paragraph{Dynamics of inequality and inclusivity in participation.} To assess the global distribution of research contributions, we computed the Gini coefficient and Pielou's evenness index on a yearly basis. When restricting the analysis to the 62 countries that have historically contributed to MICCAI, the Gini coefficient ranged between 0.86 and 0.91. However, since our goal was to assess participation inequality relative to the full set of potentially eligible countries, we instead considered all countries listed in the ISO 3166 standard \cite{iso3166}. In this setting, zero participation represents a meaningful outcome rather than missing data. Including non-participating countries therefore increased the Gini coefficient to values between 0.95 and 0.97, indicating very high degree of inequality in global participation. This reflects not only disparities among contributing countries, but also an unequal participation, with many countries lacking representation in MICCAI. Because Gini values become highly concentrated near the upper bound when all countries are included, differences become visually compressed, especially relative to Pielou's index. For clarity, we therefore present Gini trends based on the 62 countries in Fig.~\ref{fig:ginipielou}.

We conducted linear regression analyses to evaluate temporal trends in inequality (Gini index) and diversity (Pielou index). Fig.~\ref{fig:ginipielou} shows that the inequality has slowly decreased since the beginning of the conference, with a small but significant slope ($slope=-0.001, R^2=0.249, t=-2.94, p=0.007$). In contrast, Pielou's evenness shows a stronger negative trend over time ($slope=-0.004, R^2=0.62, t=-6.57, p=5.73e-07$), indicating that contributions are becoming increasingly concentrated among a smaller subset of countries (e.g., USA, China, UK, Germany, France, Canada), likely driven by strong growth from a small group of countries, particularly China. 

Analyzing simultaneous local minima/maxima of these indices helps reveal periods of higher concentration (less diversity) and greater inclusivity (more diversity). Notably, years with higher research concentrations indicating that contributions were dominated by a smaller subset of countries, align with conferences held in the United States (2005, 2014), China (2010), and France (2012). Conversely, years with increased diversity,  where research contributions were more evenly distributed across countries, coincide with conferences held in Denmark (2006), Canada (2011), Japan (2013), Germany (2015) and Singapore (2022). These findings suggest an association between geographical conference location and participation diversity, with some regions showing broader international involvement, potentially due to factors such as accessibility, attractiveness, or visa policies of the host country. However, the variation between local minima/maxima is very small, and a more detailed analysis is needed. 

\subsection{Structural position in collaboration networks}
\label{subsec:collabnetworks}
\paragraph{Research collaborations remain largely national and stratified.} 
Networks analysis and Louvain community detection reveal that research affiliations are predominantly concentrated within single-country boundaries, with relatively few extending across countries. 6,494 (66.6\%) papers involved authors from a single country, while 3,250 papers (33.4\%) featured international collaborations. Network centrality measures reveal a concentration of connectivity and influence among institutions based in wealthier economies. The top five countries by degree centrality, reflecting the number of direct collaboration ties, are all from \textit{high-income} countries, with the United States, United Kingdom, and Germany, occupying the highest positions. Similarly, betweenness centrality, which captures a country's role as a bridge within the collaboration network, further highlights the structural dominance of these countries. China is the only non-\textit{high-income} country appearing among the top ten. Average centrality measures reveal a pronounced disparity in network positions: \textit{high-income} countries exhibit an average degree centrality of 0.20, nearly double that of \textit{upper-middle-income} (0.11), four times that of \textit{lower-middle-income} (0.05), and nearly seven times that of \textit{low-income} (0.03). This difference becomes even more striking for betweenness centrality with values of 0.0263 for \textit{high-income} countries compared to 0.0036, 0.0005, and <0.0001 for the other groups. These results suggest that institutions from lower-income economies play a comparatively limited role in bridging connections, and tend to occupy more peripheral, less connected positions within the collaborative network. 

We show in Fig~\ref{fig:typologyscpvsmcp} a comparison between the number of Single-Country Paper (SCP) and Multi-Country Paper (MCP) to understand how are the collaboration relationships between the countries. \textit{Domestic-oriented} countries primarily produce research internally, suggesting strong national research capacity and limited dependence on international collaboration (e.g., Republic of Korea, Japan, Israel). \textit{International hubs} are highly collaborative countries that are well integrated into the global research network while still retain some domestic research capacity (e.g., Singapore, Australia, Denmark, Spain, Brazil, Argentina, Egypt, South Africa). \textit{Balanced} countries show a mix of domestic and international collaboration, reflecting both strong national research strategy and external engagement (e.g., USA, China, France, Canada). \textit{Exclusively-collaborative} have only MCP, fully relying on collaborations for publication, e.g., Bangladesh, Romania, Hungary, Chile, Rwanda, and Uganda. 

\begin{figure*}[t]
        \centering
        \includegraphics[width=\linewidth]{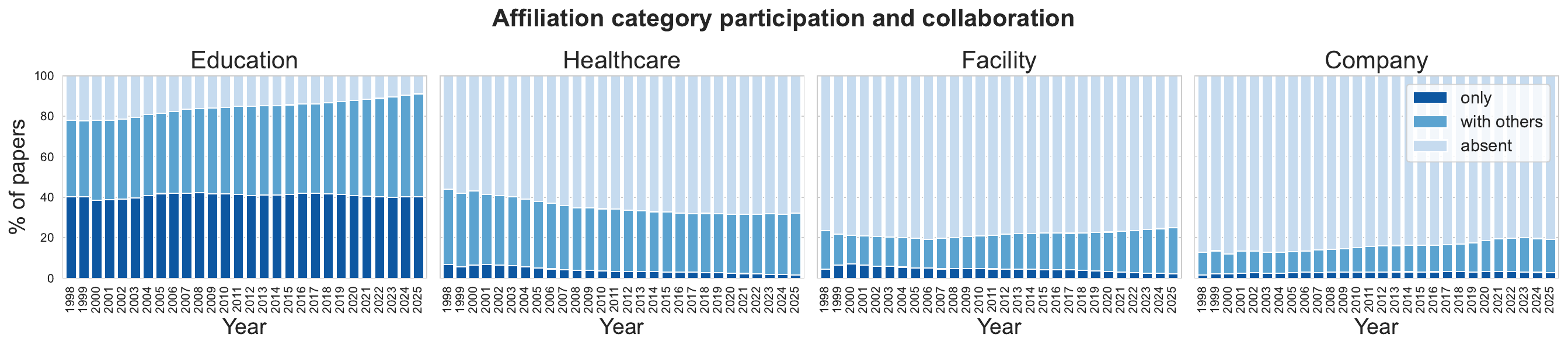}
        \caption{Educational institutions remain central to MICCAI collaborations. Percentage of papers that include each affiliation category, distinguishing between papers with that category alone (only), in combination with others (with others), and those where the category is absent (absent).}
        \label{fig:afftypebinary}
\end{figure*}

\begin{figure*}[t]
        \centering
        \includegraphics[width=0.7\linewidth]{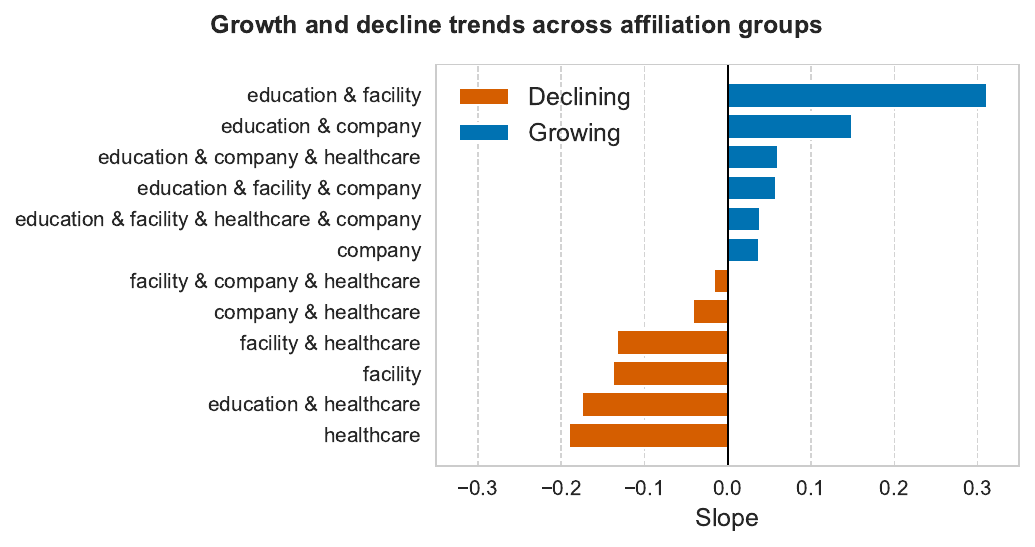}
        \caption{Academic institutions are increasingly collaborating across categories, with a growing presence of companies. Affiliation categories exhibiting declines are shown by red bars on the left, while those experiencing growth are shown by green bars on the right.}
        \label{fig:afftypetrends}
\end{figure*}

\subsection{Institutional composition and public--private collaboration} \label{subsec:institutionalcollab}

\paragraph{Educational institutions remain central to MICCAI collaborations.} Fig.~\ref{fig:afftypebinary} shows the temporal evolution of the affiliation types contributing to MICCAI. The conference remains predominantly education-driven, with approximately 40\% of papers in each edition originating exclusively from educational institutions. Moreover, between 80--90\% of papers involve at least one educational institution, often through collaborations with hospitals or medical centers. 

The affiliation type collaboration trend analysis reveals a clear increase in collaborations involving educational institutions, particularly education--facility and education--company partnerships, which exhibit the strongest positive growth over time, see Figure~\ref{fig:afftypetrends}. More complex collaboration structures, including combinations of education, company, healthcare, and facility affiliations, are also becoming increasingly common. These patterns reflect the expanding role of academic institutions as central connectors within broader collaborative ecosystems, particularly in configurations resembling public--private partnerships (PPP), facilitating knowledge transfer and real-world applications.

In contrast, collaborations centered primarily on healthcare and facility institutions show consistent declines. The strongest negative trends are observed for facility–healthcare and education–healthcare collaborations, alongside decreases in standalone healthcare and facility. Overall, these trends suggest a gradual shift in MICCAI toward broader multi-institution and industry-oriented collaborations, where PPP-like structures play an increasingly prominent role, while purely healthcare-centered participation is becoming less prominent.

Importantly, despite these declines, healthcare remains present within growing multi-institutional collaborations that include industry and academic partners, suggesting a shift in its role rather than a disappearance. Overall, these results indicate a structural transition in MICCAI toward more diversified and industry-integrated collaboration patterns, where educational institutions increasingly anchor collaboration networks, while purely healthcare- or facility-centered participation becomes less prominent unless embedded within broader, multi-actor configurations.

\subsection{Trends in research topics: agenda-setting over time}
\label{subsec:trendsInResearchTopics}
The volume of open-access papers at MICCAI has grown substantially over the last five years, nearly doubling from 531 papers in 2021 to 1,027 papers in 2025 -- an overall expansion of +93.4\%. Alongside this volume expansion, the average number of category labels per paper increased dramatically, reaching a peak of 5.41 in 2025, compared with 3.25-3.88 in previous years. Rather than indicating a sudden shift toward hyper-interdisciplinary research, this surge is an artifact of changing platform category organization. The 2025 submission site enforced a rigorous (Main Category $\rightarrow$ Sub category) hierarchical taxonomy that led to near-universal coverage ($\ge 86\%$) across four primary categories: \textit{Applications, Body, Modalities,} and \textit{Machine Learning}. In contrast, historical data from 2021-2024 reflects partial tagging, where authors frequently left papers untagged on at least one main category.

\paragraph{Core pillars vs emerging paradigms.} Despite structural shifts in tagging, core anchor topics continue to form the baseline of the conference. \textit{Image Segmentation} and \textit{Computer Aided Diagnosis} remained the top two subcategories within the \textit{Applications} category every single year. Their absolute volume grew throughout (from 320 to 608 papers), but their combined share of all \textit{Applications} labels declined steadily, from 58\% in 2021 to 45\% in 2025, as the axis widened from 10 to 12 subcategories -- indicating dilution by newly differentiated topics rather than a loss of centrality. Similarly, \textit{MRI} and \textit{CT} remained the dominant imaging modalities. Even when \textit{CT/X-Ray} overtook \textit{MRI} for the first time in 2025 (325 vs. 281 papers), this shift was largely driven by the organizational merging of the two modalities (CT and X-Ray) into a single category.

At the same time, emerging research trends reveal a clear movement toward label-efficient learning techniques and generative models. While \textit{Self-, Semi-, and Weakly-supervised Representation Learning} remained the largest overall \textit{Machine Learning} subfamily from 2021 to 2024, \textit{Foundation Models} emerged in 2024 (87 papers) to become a top-3 subcategory by 2025 (135 papers). This mirrors changes on the \textit{Applications} category, where generative themes were subcategorized into a dedicated \textit{Image Synthesis / Augmentation / Super-Resolution} subcategories in 2025, instantly becoming the third-largest subcategory with 199 papers. Methodological attention to clinical equity and model robustness continued to strengthen over time. \textit{Domain Adaptation / Generalization} consistently ranked within the top eight categories and re-entered the top five in 2025, while \textit{Algorithmic Fairness} and a dedicated \textit{Health Equity / Special Topic} cluster emerged in 2024. Interestingly, this equity cluster consolidated from 11 diverse subcategories in 2024 to 4 focused topics in 2025, primarily addressing low-cost and limited-resource settings, suggesting a maturation from broad thematic classification into well-defined clinical niches.

\paragraph{The long tail and cross-category intersections.} Evaluating the smallest, stable subcategories across the 2021-2024 taxonomies reveals a highly consistent, non-random long tail of underexplored research areas. In the \textit{Machine Learning} category, \textit{Active Learning} remained among the three smallest \textit{ML} subcategories, alongside \textit{Reinforcement Learning} and \textit{Meta-learning} (averaging $\approx 7.5$ papers/year). This output was substantially lower than that of other mid-sized topics like \textit{Uncertainty} ($\approx 38$ papers/year) and \textit{Interpretability} ($\approx 64$ papers/year). In the \textit{Modalities} category, \textit{Biophotonics} and \textit{EEG/ECG} consistently occupied the bottom slots. Within the \textit{Surgery} category, hardware-oriented subdomains like \textit{Medical Robotics} and \textit{Haptics} or \textit{Interventional Simulation Systems} were persistently less represented compared with data-driven surgical scene understanding. Finally, \textit{Fetal Imaging, Gynecology, Urology}, and \textit{Spine} represented the most structurally underrepresented anatomical regions.

\begin{figure*}[t]
        \centering
        \includegraphics[width=0.77\linewidth]{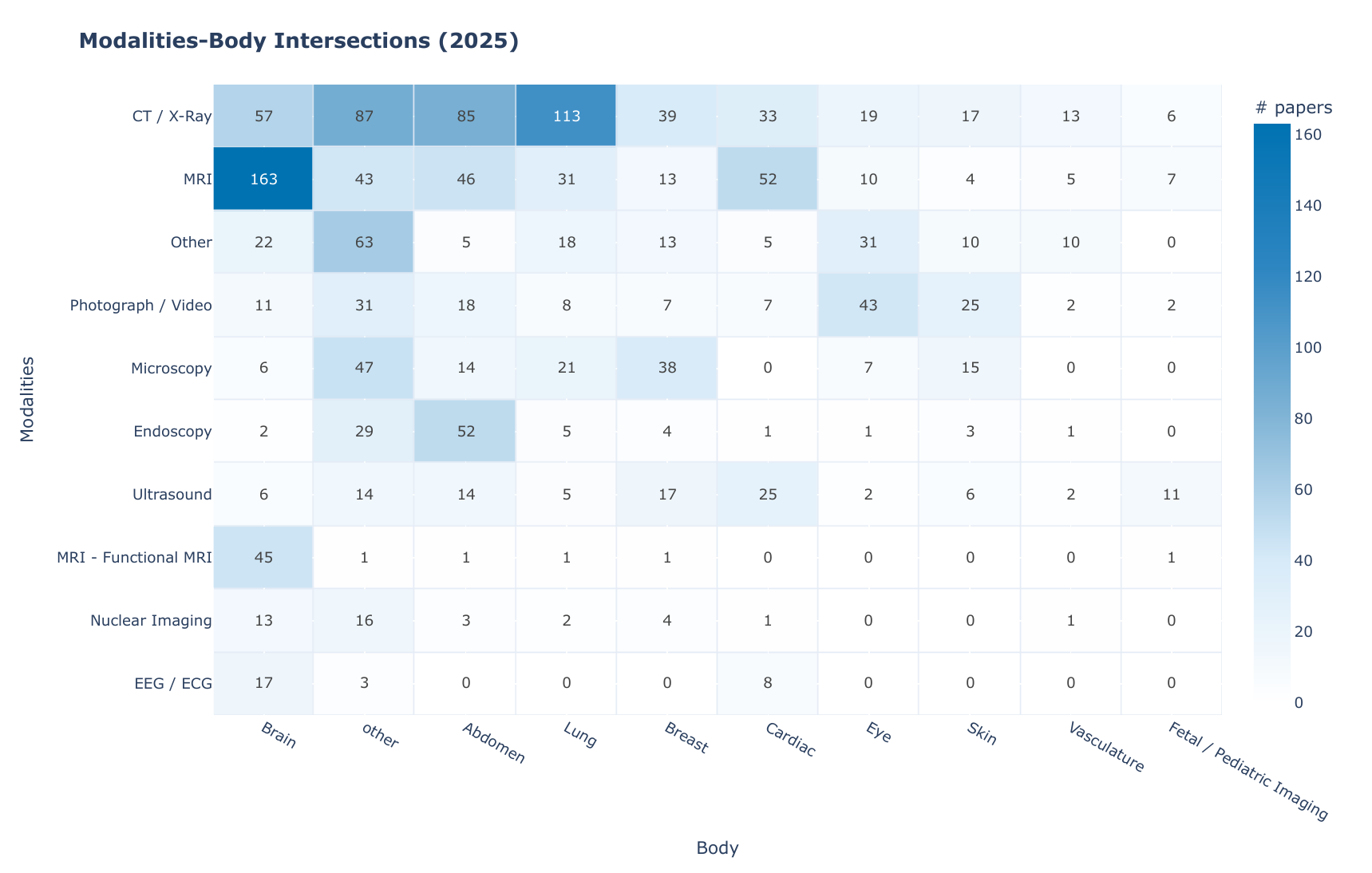}
        \caption{Cross-category intersection for papers from 2025 across the \textit{Modalities} and \textit{Body} (\textit{Clinical Applications}) categories. Cell values indicate the number of intersecting papers, with color intensity proportional to the count.}
        \label{fig:body_modalities}
\end{figure*}

Cross-category co-occurrence mapping for 2025 confirms that these categories intersect in highly structured, clinically intuitive patterns (See Figure~\ref{fig:body_modalities} \footnote{The \textit{Other} label that appears in both heatmaps is part of MICCAI's 2025 taxonomy and was offered as an explicit option in five of the six main categories. It is not an aggregation of small subcategories created by our analysis. Authors selected it frequently, thus cells in the corresponding rows and columns should therefore be read as unresolved rather than as a coherent category.}). \textit{MRI} pairs predominantly with \textit{Brain}, which at 163 papers is the strongest co-occurrence in the entire 2025 dataset once the blanket \textit{Deep Learning} tag is set aside. Brain imaging also stands out as the only anatomical region mature enough to possess its own dedicated application subcategory (\textit{Brain Network Analysis}, 57 papers).

\begin{figure*}[t]
        \centering
        \includegraphics[width=0.77\linewidth]{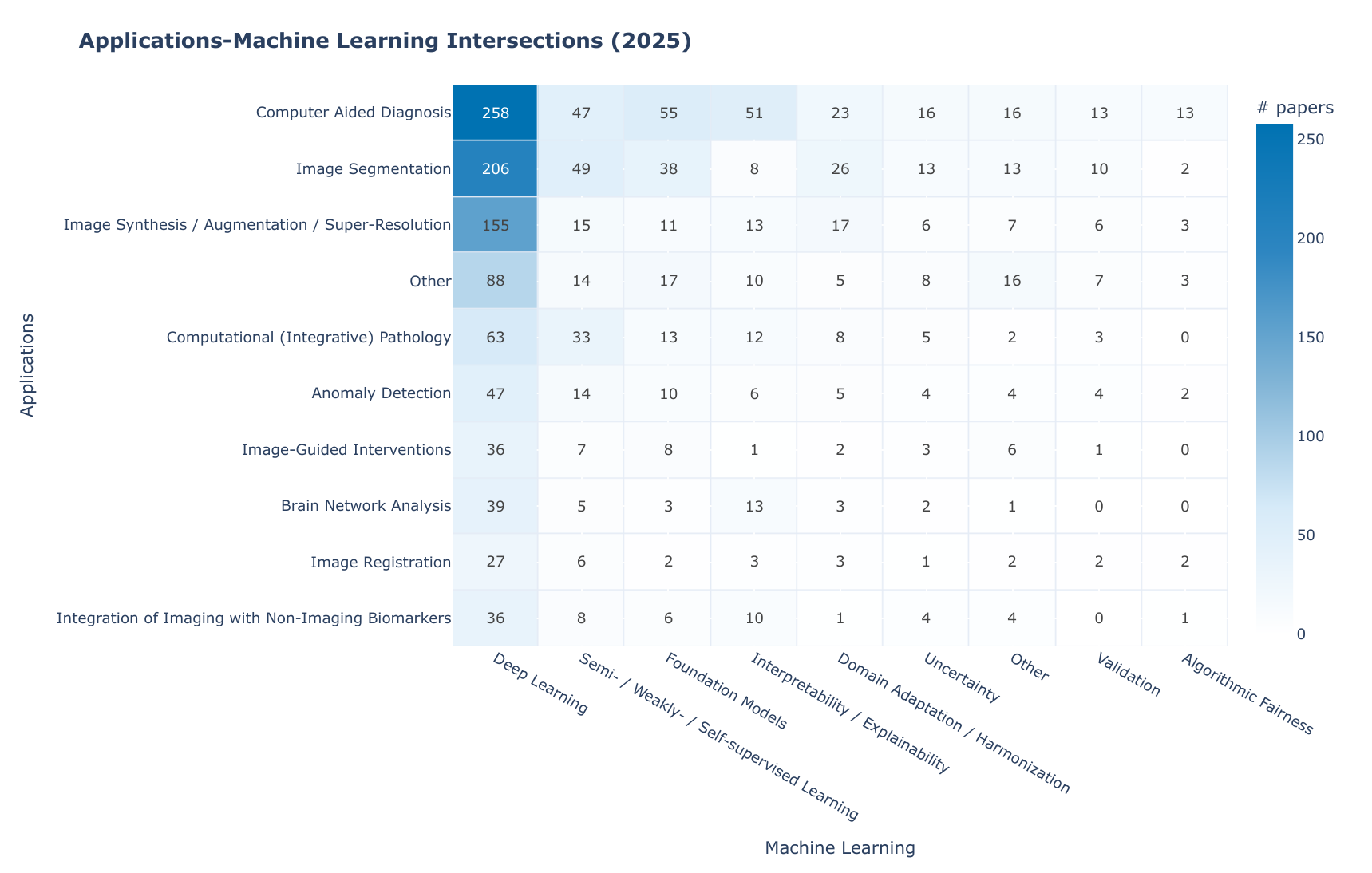}
        \caption{Cross-category intersection for papers from 2025 across the \textit{Applications} and \textit{Machine Learning} categories. Cell values indicate the number of intersecting papers, with color intensity proportional to the count.}        
        \label{fig:app_ml}
\end{figure*}

Crucially, intersecting these categories provides a more actionable definition of a ``research gap'' than evaluating individual low-volume categories. While \textit{Foundation Models} and \textit{Interpretability} heavily intersect with \textit{Computer Aided Diagnosis}, indicating that newer ML methods are validated on diagnostic tasks first, dozens of specific application-technique pairings appear only once or twice (see Figure~\ref{fig:app_ml}). These sparse intersections point to potentially unexplored frontiers where mature methodologies have not yet be applied to established clinical problems.

\section{Discussion} \label{sec:discussion} 
\paragraph{Summary of findings.} Our findings show that, although disparities have slightly decreased over time, significant geographic inequalities persist. Northern America, Europe and Asia each account for roughly 30\% of publications, while Latin America and Africa each contribute less than 1\% . Overall, nearly 50\% of all papers originate from only two countries (the USA and China), with China showing rapid growth since 2017. Collaboration network analysis reveals clear disparities: \textit{high-income} countries hold a central role of influence, while \textit{lower-middle} and \textit{low-income} countries remain peripheral and rarely act as bridges. Countries also differ in collaboration patterns, we find a group of countries that contribute to MICCAI only through partnerships with other nations. Beyond participation, the research agenda itself shows a similar concentration. A small set of core pillars, including segmentation, diagnosis, MRI, and CT, accounts for the majority of submissions every year, while dozens of application--method and organ--modality pairings appear only once or twice across five years of tagged data, revealing a persistent long tail of underexplored intersections.

Educational institutions remain the main drivers of MICCAI research, contributing each year about 40\% of papers independently and appearing as collaborator in up to 90\% of all publications. Collaboration patterns show a clear rise in partnerships between academia and other sectors, particularly with facilities and industry, alongside increasing prevalence of multi-institutional configurations resembling public–private partnerships. While broader AI and computer vision conferences are undergoing a process of narrowing, driven by increasing industrial involvement and the growing dominance of large-scale models that require substantial computational and financial resources~\cite{klinger2020narrowing,varoquaux2025hype}, our findings suggest a different trajectory in medical imaging. This suggests that, unlike many AI and CV venues, MICCAI continues to function primarily as a scientific and educational forum, rather than an industry-centered one.

\begin{figure*}[]
    \centering
    \includegraphics[width=1.\linewidth]{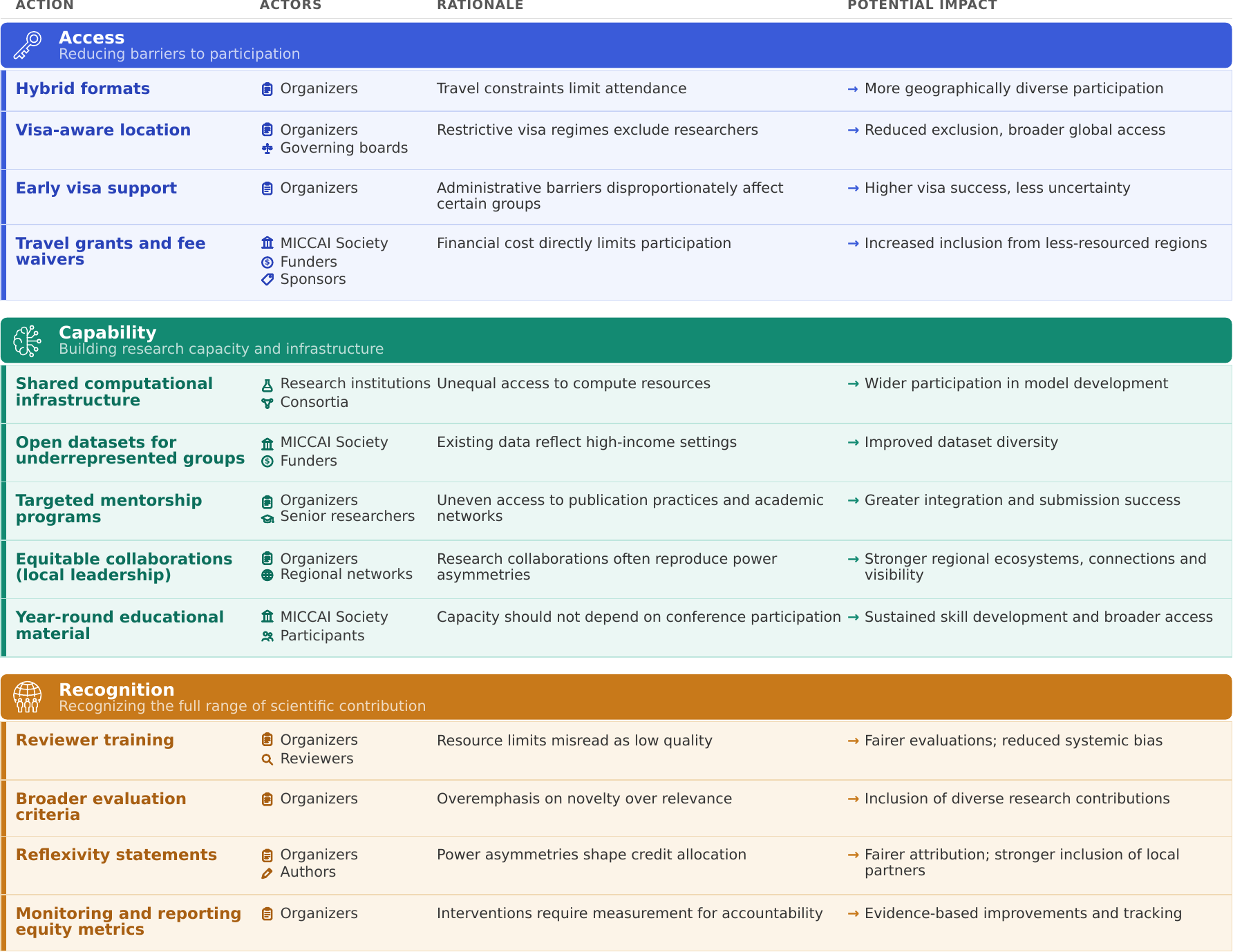}
    \caption{Actionable strategies to promote equitable participation, organized across access, capability, and recognition dimensions.}
    \label{fig:recommendations}
\end{figure*}

In the following, we discuss existing initiatives aimed at improving equity and inclusion within the MICCAI community and related research fields. We highlight current efforts to reduce geographic and economic disparities and identify remaining challenges. Building on these observations, we outline concrete strategies and recommendations to promote more balanced global participation and support underrepresented regions. These are synthesized and summarized in Figure~\ref{fig:recommendations}.

\subsection{Existing initiatives} 
The MICCAI community has introduced a range of initiatives in recent years to address disparities. 

\textbf{Geographic and regional diversity}. The \textbf{Reinforcing Inclusiveness \& Diversity and Empowering MICCAI in Low-to-Middle Income Countries (RISE) initiative}~\cite{rise2020miccai}, launched in 2020, responds to the extremely limited participation from low- and middle-income countries (LMICs), aiming to strengthen geographic diversity, support minority researchers in LMICs, and build a strong network both within and across regions to reduce global disparities in health and imaging. \textbf{The AFRICAI network}~\cite{africai2023}, launched during MICCAI 2022, represents a growing effort to strengthen the medical imaging AI community across Africa. Since its inception, AFRICAI has focused on community engagement and knowledge exchange, including the organization of the first AFRICAI Summer School on Medical Imaging in 2023. Its activities expanded with the ``\textbf{MICCAI meets Africa}'' Workshop (2024)~\cite{miccaimeetsafrica2024}, providing a platform to highlight research and challenges specific to the region. AFRICAI was formalized as a MICCAI Special Interest Group (SIG) in 2025, gaining official recognition by the MICCAI Society.

\textbf{Financial support mechanisms} have also expanded to promote inclusion. The MICCAI Society awards travel and conference grants, which since 2025 have incorporated broader diversity criteria, including not only gender, race and ability, but also geographic representation-specifically targeting researchers from LMICs and those with historically low representation in MICCAI. Additional support is provided through initiatives like RISE and dedicated challenges, which offer targeted travel awards. 

\textbf{Mentorship and educational material}. The \textbf{MICCAI Society Mentorship Program} \cite{miccai_mentorship} (running since 2018) matches mentees to mentors on the basis of research area, geography, and goals, with a stated focus on under-represented collectives and a membership-fee exception for applicants from LMICs. The associated \textbf{Young Researcher Excellence Training Program (YRETP)} \cite{miccai_yretp} offers online, small-group guidance through the full paper-preparation process (from both author and reviewer perspectives) and has explicitly prioritized regions historically under-represented at MICCAI. \textbf{RISE-MICCAI} has recently launched a \textbf{Publication Accelerator} \cite{rise_miccai_accelerator} that pairs first-time MICCAI lead authors, with priority given to underrepresented groups and applicants from LMICs, with a senior research advisor for one-on-one guidance on scoping, experiments, and manuscript preparation ahead of submission. Beyond mentoring, the MICCAI Student Board (MSB) also runs the \textbf{MICCAI Educational Challenge }\cite{msbeducational} to create and share educational materials, covering fundamental concepts in medical image computing and computer-assisted interventions.

\textbf{Thematic initiatives} around responsible AI have been introduced to broaden the diversity of topics. In this context, fairness and bias in AI have gained space. The Fairness of AI in Medical Imaging (\textbf{FAIMI}) Workshop~\cite{faimi2022miccai}, first held in 2022 as a virtual event, has since become a recurring part of MICCAI and is also a SIG group officially recognized by the MICCAI Society. Its evolution, including integration into broader ethics-focused workshops, reflects growing engagement with responsible and equitable AI practices within the community. Building on this momentum, the ``\textbf{Health Equity Track}'' was introduced alongside the first MICCAI conference held in Africa in 2024. It established a dedicated topic on ``Equity in medical imaging, including accessible and scalable solutions,'' which has been maintained in subsequent editions of the conference. Research in resource-constrained settings has also gained visibility, particularly with a focus on LMICs since the introduction of the \textbf{MIRASOL} Workshop (2025)~\cite{mirasol2025miccai}. In parallel, the Open Data Initiative (2024)~\cite{opendata2024miccai} promotes the development of datasets for underrepresented populations and diseases, addressing critical shortcomings in data availability. These efforts are complemented by the MICCAI Society Award for the Advancement of Health Equity (2026) and Open Data Micro-Grants, which support data curation, annotation, quality control, and public release—highlighting both a commitment to equity and the need to better recognize data work \cite{sambasivan2021datawork}.

\textbf{Multiple dimensions of a scientist's identity.} The \textbf{Women in MICCAI (WiM)} movement was initiated at MICCAI 2015 and ISBI 2016, with its board formally established in 2016. Its primary aim is to increase female participation in the field and ensure fair and equitable career promotion for women researchers and faculty. As a secondary, exploratory analysis, we examined author-level patterns using name-based sex inference via the Wiki-Gendersort~\cite{berube2020wiki} library (see details in Appendix~\ref{subsec:sexinference}). We note an important limitation, since these tools infer sex assigned at birth rather than self-identified gender, and their accuracy varies across cultural and linguistic contexts, performing less reliably for several Asian and other underrepresented naming traditions than for Western names \cite{santamaria2018comparison,vanhelene2024inferring}. We therefore treat this analysis as indicative rather than definitive. Our data show that male authorship dominance has declined over time from approximately 95\% to 80\%, yet male overrepresentation remains more pronounced at the last-author position than the first-author position. These patterns align with prior findings: a recent examination of AI and Critical Care literature between 2010 and 2022 reported that while over 70\% of publications included at least one inferred female author, fewer than 25\% had a woman as first or last author \cite{charpignon2024diversity}. 

These gender disparities likely intersect with geographic patterns, but we do not examine their joint effects. Name-based inference is less reliable for naming traditions underrepresented in training data \cite{santamaria2018comparison,vanhelene2024inferring}, often corresponding to regions that are also underrepresented in our geographic analysis. As a result, intersectional analyses would concentrate measurement error in the very groups they seek to characterize (\textit{e.g.}, women researchers in LMICs). We consider geography-gender intersectionality better suited to survey- or interview-based methods. WiM is part of a broader ecosystem of affinity groups across AI, such as Queer in AI~\cite{queerinai2017}, Black in AI~\cite{blackinai2017}, and LatinX in AI~\cite{latinxinai2018}, that work to foster inclusion for a broader range of underrepresented communities. Other dimensions of participants' diversity merit investigation through qualitative methods, but lie beyond the scope of this study and represent an important direction for future work. 

\paragraph{Assessing initiatives beyond intentions.} Interventions can only be evaluated if their effects are tracked. We recommend that the community, using the kind of longitudinal, publicly auditable bibliometric record assembled in this study, monitor and report participation, collaboration, and agenda-setting metrics across editions, so that initiatives such as RISE, the Health Equity Track, and satellite venues can be assessed against measured outcomes rather than intentions. Therefore, moving from representativeness as a metric, counting who is present, to empowerment as a condition means ensuring that a broad range of researchers can produce knowledge that the field receives, recognizes, and builds upon. While all the strategies outlined above can address structural barriers at the conference level; researchers who may not be well informed, do not have network access, institutional backing, and familiarity with submission culture, may still be unintentionally excluded. Addressing these sources of exclusion requires investment in local research communities over time, not only in individual conference attendance. Following the decolonizing global health movement, support should move beyond one-off travel grants toward sustained funding streams that build local research communities over time \cite{abimbola2020decolonising}. 

\subsection{Actionable strategies for equitable participation}
The concentration of contributions in high-income countries, evident in our findings, is a reflection of where research capacity exists and it is allowed to develop. Many researchers whose work might be shaped by resource constraints rather than scientific weakness could be systematically discouraged to participate and be at a weaker position to get accepted in MICCAI or similar conferences.

This is an ethical concern: knowledge produced under unequal conditions could be developing AI models aligned primarily with high-income populations, disease burdens, and clinical workflows. Models trained on large datasets from well-resourced institutions, evaluated in high-income contexts, and guided by narrow priorities become positioned as universal \cite{celi2022sourcesofbias}. The geographic and institutional inequalities we observe are also a scientific challenge, we would expect this to limit the quality, generalizability, and clinical relevance of the field. As Kuhlman~\etal~argue, there is no computation without representation, because the composition of a research community is inseparable from the quality of the research it produces~\cite{Kuhlman2020Representation}.

Addressing this requires moving beyond awareness to structural interventions that target the specific mechanisms of exclusion. Evidence from related research communities (global health, AI ethics, and human-computer interactions), offer concrete examples on how others are addressing these challenges. Here, we propose actions at multiple levels to increase global representativeness of researchers, acknowledging different capacities and increasing visibility of results that might be outside the norms of the field. 

\paragraph{(1) Reducing barriers to participation.} The most direct form of exclusion from a scientific community is physical: visa restrictions, travel costs, and registration fees can prevent attendance regardless of scientific merit. For instance, at NeurIPS~2018 in Montreal, about half of the $\sim$200 African researchers invited through the Black in AI workshop~\cite{blackinai2017} were denied Canadian visas. Similar issues recurred in 2019, with many refusals reversed only after intervention by prominent organizers \cite{else2018visa,globeandmail2019visa}.

While visa decisions lie beyond conference organizers' control, host location shapes accessibility. This has prompted shifts toward hybrid and remote formats. Conferences such as FAccT, CSCW, and ISMIR now treat distributed participation as standard. A comparative analysis across multiple scientific conferences found that virtual formats increased geographic diversity and early career participation while lowering costs \cite{skiles2022virtual}. Multi-site models extend this approach: NeurIPS~2026 combines a main meeting in Sydney (Australia) with official satellite sites in Paris (France) and Atlanta (USA), allowing paper presentations without registering for the main venue \cite{neurips2026satellites}. Such formats should also be designed for low-bandwidth and asynchronous access, and be paired with stronger virtual networking and poster sessions, still the weakest aspects of online participation~\cite{skiles2022virtual}.

Host selection remains relevant where in-person participation matters. There is growing support for rotating locations or choosing visa-accessible venues. ICLR 2020 moved to Addis Ababa in response to the Canadian visa refusals \cite{globeandmail2019visa}, and MICCAI's 2024 edition in Morocco reflects a similar effort. Conference organizers can further ease access through early invitation letters, supplying attendee lists to immigration authorities in advance, and providing dedicated visa-support contacts.

Financial barriers also persist: researchers may be able to travel but be unable to afford registration. Initiatives like RISE and the MICCAI Society provide travel grants that help address this; scaling them and clarifying eligibility criteria could broaden impact. Combining fee support with hybrid participation appears more effective than either measure alone \cite{skiles2022virtual}.

\paragraph{(2) Building research capacity and infrastructure.} The rise of \textit{Foundation Models} as an emerging paradigm (Section~\ref{subsec:trendsInResearchTopics}) illustrates how existing resource disparities risk becoming methodological barriers as the field becomes more dependent on large-scale compute and data. Unequal access to compute infrastructure and large annotated datasets cannot be solved at the individual level or through diversity statements alone. Collaborative frameworks that share resources across regions can help address these challenges. Distributed ML approaches change the collaboration paradigm because rather than exporting tools developed elsewhere for local validation, it centers local researchers as primary knowledge-makers from the earliest stages of problem definition. For example, federated learning enables contributions in resource-constrained settings without centralized data or equal compute, turning site diversity into a scientific advantage \cite{cheung2023methods, rieke2020future}. 

Benchmark reform is equally important. Requiring evaluation across multiple geographic and demographic contexts, and rewarding clinical relevance alongside methodological novelty, would reshape what the field recognizes as rigorous science. Recent work highlights biases in medical imaging AI challenges, they show substantial biases in dataset composition, including geographic, modality-, and task-related imbalances, showing that current benchmarks fail to reflect real-world clinical diversity~\cite{reinke2025medical}. The BIAS initiative \cite{maierhein2020bias}, endorsed by MICCAI, which currently focuses on standardizing challenge reporting and metric selection, could in principle extend its scope to include geographic and demographic representativeness of datasets as explicit quality criteria, alongside clinical validity and fairness metrics. Similarly, MICCAI's Open Data Initiative \cite{opendata2024miccai} and its micro-grants support more inclusive data curation from underrepresented populations and diseases. 

Capacity is also built through people, not only infrastructure. We recommend scaling existing mentorships programs such as the MICCAI Mentorship \cite{miccai_mentorship} and the Young Researcher Excellence Training Program \cite{miccai_yretp}, with a focus on under-represented regions. These efforts could be further strengthened through partnerships with regional organizations and venues that already convene the relevant communities, such as the Deep Learning Indaba, an annual pan-African gathering rotating across locations including Tunis, Dakar, and Kigali, with IndabaX satellites in over 20 countries, supporting co-located tracks, shared calls, and joint workshops~\cite{indaba}. Such collaborations, together with partnerships with clinical and academic actors in target settings, can help center local researchers as primary knowledge-makers. We also recommend improving access to educational resources. Expanding low-bandwidth, year-round training resources could help broaden participation, this could be combined with the educational material on the MICCAI website or the MICCAI Educational Challenge \cite{msbeducational} by the MSB. Overall, our recommendation is not to create new structures, but to scale and resource existing programs, targeting them deliberately at the lowest-represented regions. 
  
\paragraph{(3) Recognizing the full range of scientific contribution.} The dominance of English, adherence to international submission conventions, and the structural privileging of methodological novelty over clinical validation reflect historical norms, rather than inherent features of scientific rigor. This privileging is itself a values-laden judgment, not a neutral technical fact: even choices as narrow as model architecture encode underlying research values rather than purely objective criteria \cite{baxter2025exploring}, and the same is true of the field's broader standards for what counts as a competitive submission. As such, they can, and should, be reconsidered. Reviewer training that explicitly distinguishes between methodologically \emph{weak} work and methodologically \emph{constrained} work, as developed in social computing and fairness research, could help ensure that carefully curated small datasets, locally relevant clinical problems, and studies conducted under computational constraints are not unfairly penalized as proxies for low quality. Treating such contributions as second-tier outputs structurally disadvantages the researchers most likely to produce them and narrows the range of knowledge the field is willing to build upon. Similarly, work on dataset curation \cite{sambasivan2021datawork} and clinical validation is often undervalued, despite deserving recognition as primary scientific contributions in their own right. Our analysis of recent MICCAI years show that the majority of papers focus on methodological contributions, while only a small fraction address validation and algorithmic fairness (see Fig.~\ref{fig:app_ml}). It was only in 2024, with the introduction of the Health Equity Track alongside the first MICCAI conference held in Africa, that a step was taken toward recognizing equity-focused research. 

These structural patterns also shape who is able to lead and define research agendas. The ``stuck-in-the-middle'' phenomenon, where researchers from under-represented regions and women are less likely to occupy first or last authorship positions that shape the conceptual agenda, has been documented in global health \cite{hedt2019stuck} and medical imaging research \cite{charpignon2024diversity,lee2025representation}. Authorship equity frameworks developed in global health, including reflexivity statements that require corresponding authors to account for the distribution of analytical labor and the role of local partners from study inception, could be effectively adapted to MICCAI's international collaborative submissions.

\paragraph{Broadening the scientific agenda.} MICCAI's research agenda, like its participation base, is unevenly distributed: a small set of core pillars (segmentation, diagnosis, MRI/CT) accounts for the large majority of submissions each year, while dozens of application--method and organ--modality pairings appear only once or twice across five years of tagged data. This concentration is not necessarily a sign of low scientific value in the long tail; it may instead reflect the areas where established datasets, benchmarks, and reviewer familiarity already exist, resources that are themselves unevenly distributed across the community. Encouraging submissions, challenges, or dedicated tracks around these underexplored intersections could therefore serve two goals at once: expanding methodological coverage into clinically relevant but understudied areas, and creating entry points for researchers, often those outside the best-resourced institutions, who work on problems that fall outside the field's current core. We also note that the category taxonomy itself has shifted substantially across editions (Section~\ref{subsec:trendsInResearchTopics}), limiting the community's ability to track these patterns consistently over time. A more stable, versioned taxonomy would make the long tail easier to monitor and address as a recurring feature of the conference, rather than an artifact we can only reconstruct retrospectively.

\paragraph{Structural determinants beyond the conference.} While the strategies above address barriers that conference organizers can influence, they operate within broader structural determinants that also shape who can participate. These interventions mitigate exclusion, but cannot replace country-level initiatives that determine research capacity. China's trajectory at MICCAI is an example of what coordinated, large-scale investment can achieve. Within a decade, China moved from marginal presence to co-dominance coinciding with national policies that expanded funding, strengthened research infrastructure, and supported researchers \cite{jiang202220}. The Healthy China 2030 vision \cite{tan2017healthy} explicitly prioritized health innovation, equity, and delivery which are conditions that enabled researchers to compete at venues like MICCAI as a critical mass. This trajectory suggests that the determinants of conference representativeness lie largely outside what any conference can control: travel grants and hybrid formats address proximate barriers, but they cannot substitute for the funding ecosystems and institutional structures that country-level policy creates. MICCAI's own initiatives are meaningful and necessary, but they operate downstream of conditions they cannot themselves generate. This example is not intended as an endorsement of a particular national model, but rather to illustrate how sustained, large-scale investment and policy alignment can shape participation in international research communities.

This does not reduce the need for action, but clarifies its scope. The under-representation documented in our findings is not unique to MICCAI because it reflects broader patterns across the field. For example, women remained underrepresented in leading authorship positions despite increasing participation \cite{charpignon2024diversity,lee2025representation}. The persistence of such inequalities even when there is institutional effort suggests that conference-level action, however well-designed, needs to be accompanied by external structural changes: in funding bodies, in hiring practices, in how research capacity is built and recognized at the national and regional level.

MICCAI is therefore well-placed not only to implement these recommendations, but also to advocate for the broader conditions that would make them effective. These challenges are not unique to this community; they are features of how the global research system distributes resources and recognition. Addressing them requires both internal reform and collective pressure on the structures beyond any single conference's reach.

\paragraph{Limitations and future work.} We acknowledge several limitations of our study and outline directions for future work. First, we used the affiliation of the first author as a proxy for regional representation, which may not fully capture the geographic diversity of all contributors. Although, we partially mitigated this by examining collaboration patterns across affiliations within each paper, future work could complement these analyses with surveys or participant interviews to obtain a more nuanced and qualitative understanding of author representation.

Second, our affiliation category analysis relies in part on manually curated affiliation annotations, which may introduce bias. In particular, affiliations were sometimes assigned to a dominant or ``main'' institution type, most often educational institutions, potentially overlooking more specific units such as schools of medicine or clinical departments. This may partly explain the relatively low representation of healthcare categories in our results. Although we attempted to standardize annotations, affiliation reporting in MICCAI submissions is not uniform. Authors may list multiple affiliations either as separate entries or combined within a single record, as exemplified in Section~\ref{subsec:datacollection}, leading to inconsistencies that could affect categorization and subsequent analysis. Future work could address these limitations by developing more systematic and automated approaches to affiliation parsing and classification, improving consistency and scalability.

Third, our analysis was restricted to main conference papers and did not include challenge papers. Because our dataset includes only accepted papers, we can identify disparities among researchers who published their work. However, we cannot distinguish whether these disparities arise from pre-submission barriers (e.g., interpretive gaps), reviewer-stage discounting (e.g., credibility deficits), or both. The bibliometric record captures only their combined effect. We focused exclusively on the MICCAI conference, analyzing over 9,000 papers, and did not consider MIDL, as one of our primary objectives was to examine long-term temporal trends, and MIDL is a relatively recent venue. Similarly we did not include journals publications (e.g., \textit{Medical Image Analysis}), as conferences typically provide greater visibility for presenting authors and more opportunities for networking. Finally, although MICCAI serves as a central and highly active hub for the community, reflected in initiatives such as RISE, Women in MICCAI, and MICCAI meets Africa, future work could extend our analysis to include these additional venues and publication types to assess whether the observed patterns generalize across the broader medical image analysis landscape.

Finally, inspired by efforts from data-centric research communities, such as MIT Critical Data~\cite{mccullum2025power}, future extensions of this work could move beyond geographic and institutional analyses to explore additional dimensions of representation, influence, and power. Incorporating bibliometric and other network analysis techniques, such as citation patterns, could help uncover how influence and knowledge flow across regions and institutions. Furthermore, analyzing acknowledgments and funding sources would provide insight into the structural factors shaping research production, collaboration, and visibility, enabling a more comprehensive understanding of equity and participation in the field.

\textbf{Researcher positionality.} 
We are a group of researchers based at Western institutions, but with backgrounds spanning Europe, South America, Africa and Asia. Our expertise covers computer science, medicine and ethics. Some authors are or have been board members of the MICCAI Student Board, Open Data Initiative, Women in MICCAI and the MICCAI Society Board. This proximity provides valuable contextual insight, but may also shape how we interpret patterns of participation, inclusion and institutional dynamics. Despite this familiarity, we faced challenges accessing structured information on special tracks and community initiatives, pointing to limitations in transparency and documentation. Our work is motivated by a commitment to advancing equity and representation. Our positionality, as researchers who are both part of the community and shaped by diverse backgrounds, informs how we interpret the findings and how we operationalize concepts such as inclusion, participation, and fairness within the social context of the field. We offer this analysis as a collaborative contribution to the community's ongoing efforts toward greater inclusion and impact. 

\section{Conclusions} \label{sec:conclusions}
Over a 28-year analysis of MICCAI contributions, we have characterized the geographical and institutional structure of the community, and introduced the Atlas of MICCAI dataset and an interactive dashboard to support ongoing, transparent monitoring of participation. Our findings reveal persistent geographic disparities, despite a slight reduction over time, with contributions remaining concentrated in a limited number of countries. Analysis of global collaboration patterns and degree centrality further underscores the dominant role of \textit{high-income} economies within the network. While MICCAI continues to be largely driven by academic institutions, our results suggest a structural shift toward more diversified and industry-integrated collaboration patterns, with educational institutions still serving as central anchors. Beyond who participates, we show that the field's research agenda is similarly concentrated. A small set of core pillars anchors the majority of work each year, while a stable, non-random long tail of underexplored topics and sparse application--method intersections point to actionable research gaps.

Just as research values shape technical decisions in medical image analysis, our findings show that they also shape the communities that get to make those decisions in the first place -- who publishes, who collaborates with whom, and whose clinical problems define the field's agenda. Recognizing this is only a first step. Our goal with this work is not to rank productivity by publication counts, but to make visible aspects of participation and agenda-setting that are often overlooked, and to translate them into action. We proposed concrete, multi-level strategies to reduce barriers to physical and financial participation, build research capacity and infrastructure, and recognize a fuller range of scientific contributions, including data curation and clinical validation. Accountability and transparency are key to advancing inclusion and measuring representativeness: understanding \textit{who} constitutes the MICCAI community today, and \textit{what} it works on, is a crucial step toward a more representative, and ultimately more rigorous, global research community.

\section*{Acknowledgments}
This work received funding from the European Union’s Horizon Europe research and innovation programme under Grant Agreement No. 101213369 (DVPS project). KK holds a Ramon y Cajal fellowship with reference number RYC2024-048256-I funded by the Spanish State Research Agency. Thanks to Juan Sebasti\'{a}n G\'{o}mez-Ca\~{n}\'{o}n for early discussions on their work, and providing the auditable framework that inspired this work. Thanks to Veronika Cheplygina for encouraging meta-research and for giving visibility to research on representativity.

\section*{Author Contributions}
Contributions follow the CRediT author statement.

\noindent \textbf{Conceptualization, Methodology}: Amelia \JimSan.

\noindent \textbf{Data Curation, Visualization, Software}: Amelia \JimSan.

\noindent \textbf{Data Analysis}: Amelia \JimSan, Kaisar Kushibar, Joan Perramon-Lluss\`{a}.

\noindent  \textbf{Writing - Original Draft}: Amelia \JimSan, Laura Arbelaez Ossa, Tanya Akumu, Kaisar Kushibar, Joan Perramon-Lluss\`{a}.

\noindent  \textbf{Writing - Reviewing}: Amelia \JimSan, Laura Arbelaez Ossa, Tanya Akumu, Vien Ngoc Dang, Kaisar Kushibar, Joan Perramon-Lluss\`{a}, Karim Lekadir, Julia A. Schnabel, Marius George Linguraru.

\noindent  \textbf{Writing - Final Draft}: Amelia \JimSan.

\noindent \textbf{Funding Acquisition}: Karim Lekadir.

\section*{Declaration of generative AI and AI-assisted technologies in the manuscript preparation process.}
During the preparation of this work, the author(s) used the Llama-4-Scout-17B-16E-Instruct model to generate weak affiliation annotations, which were later manually revised by one of the author, as described in Section~\ref{subsec:datacollection}. We include the prompt used in Appendix~\ref{subsec:prompt}. The authors also used Claude Pro and Copilot Pro to assist in generating figures, and solely for copy-editing and language refinement. We confirm that the content of this manuscript is our own.

\bibliographystyle{plain}
\bibliography{refs_amelia}

\clearpage
\appendix

\section{Supplementary Material} 

\subsection{Additional results and visualization}
Table~\ref{tab:miccai_year} provides a quick overview of MICCAI conferences across different years, summarizing key characteristics such as their locations, the number of accepted paper, and the diversity of international participation. It illustrated how MICCAI has evolved over time in terms of scale and geographical representation. 
\newcolumntype{L}[1]{>{\raggedright\arraybackslash}p{#1}}
\newcolumntype{C}[1]{>{\centering\arraybackslash}p{#1}}

\begin{table}[h]
\centering
\caption{Overview of MICCAI conferences, including location, number of papers, and number of participating countries.}
\label{tab:miccai_year}
\small
\begin{tabular}{L{0.1\columnwidth}L{0.2\columnwidth}C{0.1\columnwidth}C{0.1\columnwidth}
}
Year & Location & \#papers & \#countries \\ \hline
1998 & USA & 133 & 14 \\
1999 & UK & 133 & 18 \\
2000 & USA & 133 & 18 \\
2001 & The Netherlands & 258 & 19 \\
2002 & Japan & 183 & 18 \\
2009 & Canada & 236 & 18 \\
2004 & France & 268	& 26 \\
2005 & USA & 237 & 22 \\
2006 & Denmark & 232 & 25 \\
2007 & Australia & 237 & 25 \\
2008 & USA & 258 & 25 \\
2009 & UK & 259 & 22 \\
2010 & China & 251 & 21 \\
2011 & Canada & 249 & 21 \\
2012 & France & 252 & 21 \\
2013 & Japan & 262 & 21 \\
2014 & USA & 253 & 23 \\
2015 & Germany & 263 & 28 \\
2016 & Greece & 228 & 25 \\
2017 & Canada & 255 & 17 \\
2018 & Spain & 372 & 27 \\
2019 & China & 538 & 27 \\
2020 & Peru (o) & 542 & 34 \\
2021 & France (o) & 531 & 32 \\
2022 & Singapore & 573 & 34 \\
2023 & Canada & 730 & 32 \\
2024 & Morocco & 856 & 42 \\
2025 & Republic of Korea & 1027 & 42 \\ 
\hline
\end{tabular}
\end{table}

\begin{figure*}[t]
    \centering

    \begin{subfigure}[b]{0.95\linewidth}
        \centering
        \includegraphics[width=\linewidth]{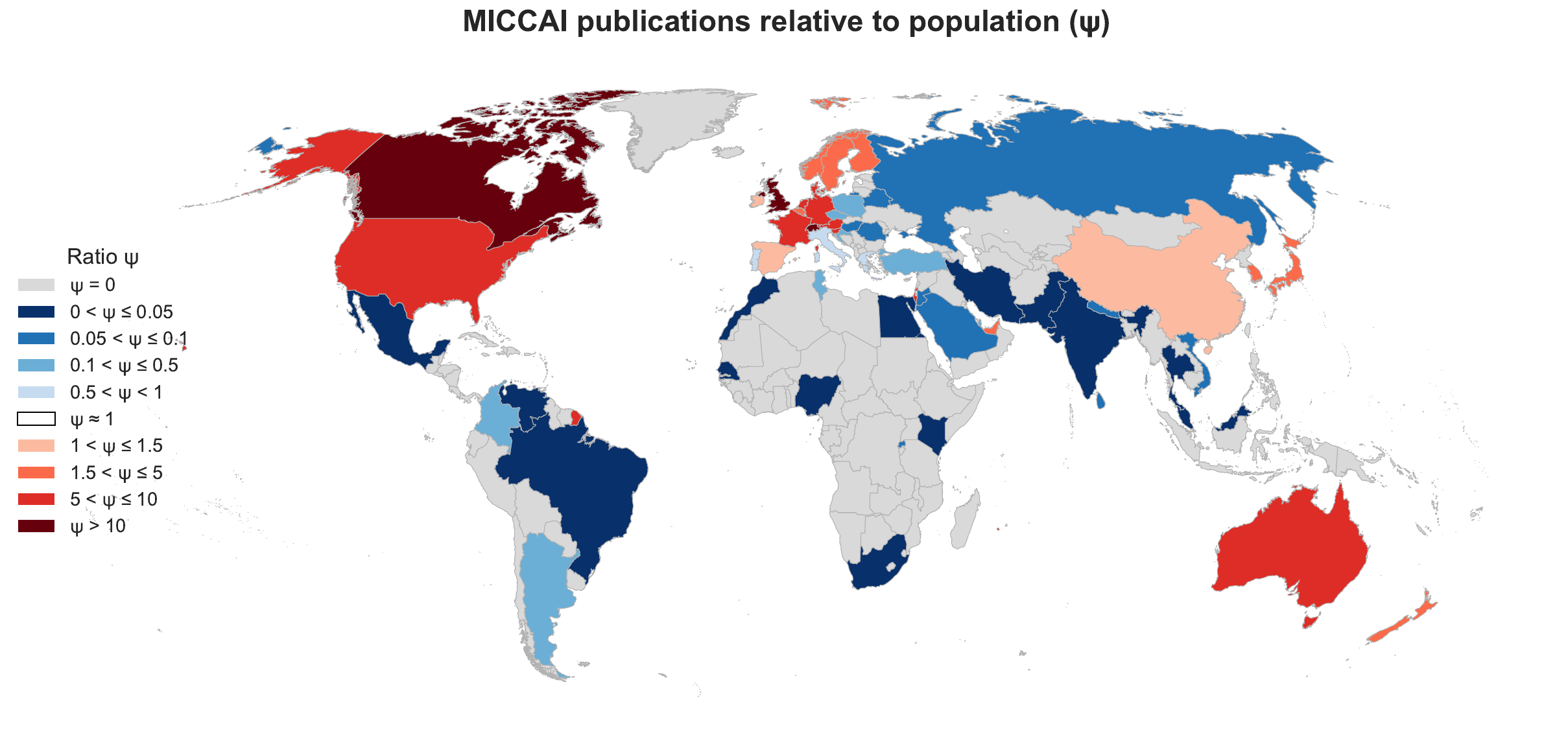}
        \caption{}
        \label{fig:worldmap-population}
    \end{subfigure}

    \vspace{0.5cm}

    \begin{subfigure}[b]{0.95\linewidth}
        \centering
        \includegraphics[width=\linewidth]{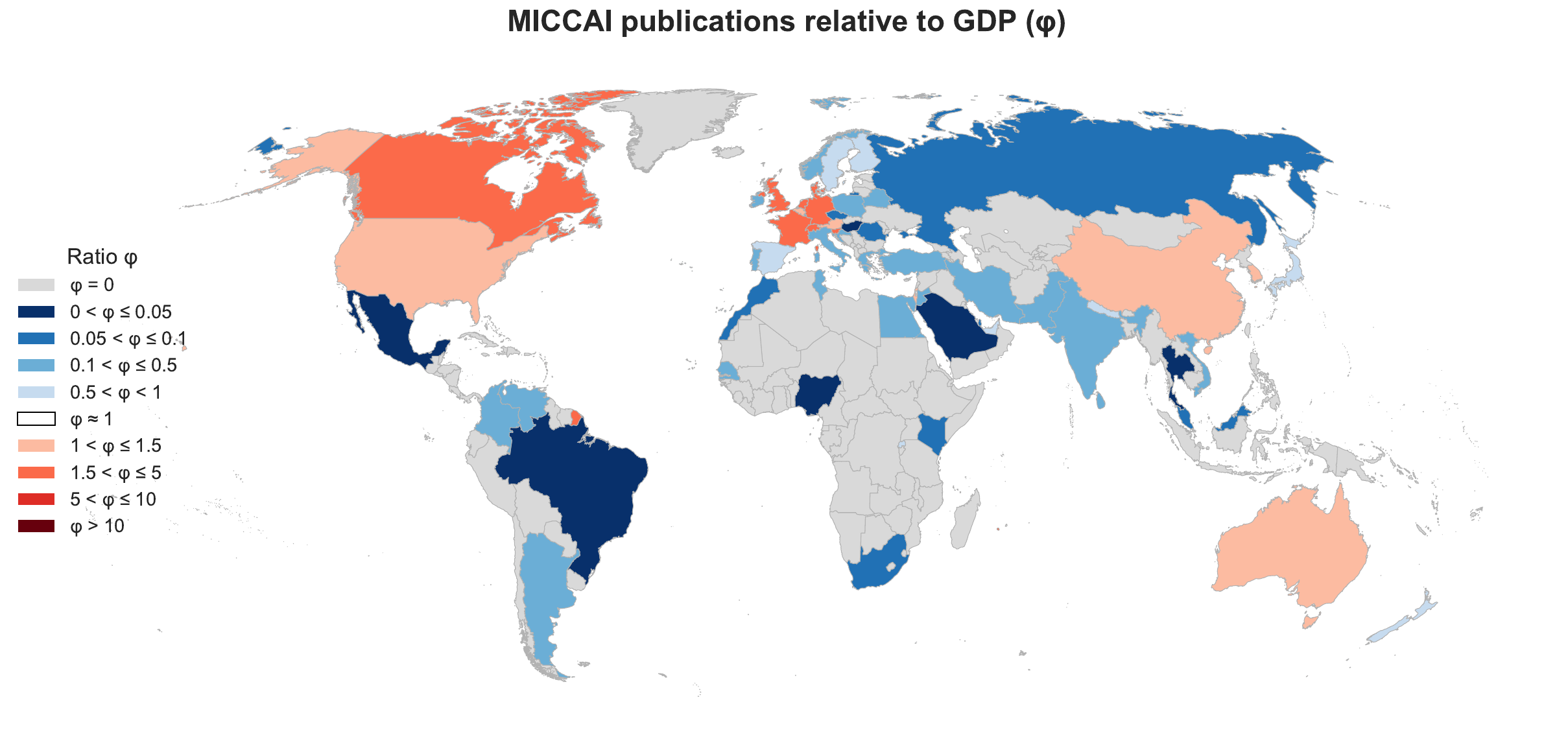}
        \caption{}
        \label{fig:worldmap-gdp}
    \end{subfigure}

    \caption{Country-level representation of MICCAI publications normalized by population and economic output. \textbf{(a)} Population-normalized publication ratio ($\psi$), defined as the share of first-author papers divided by the country's share of the global population. \textbf{(b)} GDP-normalized publication ratio ($\phi$), defined as the share of first-author papers divided by the country's share of global GDP. In both maps, values equal to 1 (shown in white) indicate publication output proportional to the reference quantity, values greater than 1 indicate overrepresentation, and values below 1 indicate underrepresentation. Countries without publications are shown in gray.}
    \label{fig:worldmap-normalized}
\end{figure*}

\textbf{Population- and GDP-normalized contribution ratios per country ($\psi$,$\phi$).} Country-level population and gross domestic product (GDP) statistics were obtained from the World Bank Open Data \cite{worldbank}. To account for differences in country size, we computed two normalized publication ratios for each country $c$: a population-normalized ratio ($\psi_c$) and a GDP-normalized ratio ($\phi_c$), based on the formulation in \cite{septiandri2023weird}. Each is defined as the country's share of all publications divided by its share of the corresponding global quantity (population or GDP):

\begin{align}
\psi_c &= \frac{\left(P_c / P_{\mathrm{total}}\right)}
               {\left(N_c / N_{\mathrm{world}}\right)}
\label{eq:psi} \\
\phi_c &= \frac{\left(P_c / P_{\mathrm{total}}\right)}
               {\left(GDP_c / GDP_{\mathrm{world}}\right)}
\label{eq:phi}
\end{align}

where $P_c$ is the number of publications associated with country $c$, $P_{\mathrm{total}}$ is the total number of publications in MICCAI, $N_c$ and $N_{\mathrm{world}}$
are the population of country $c$ and the world, and $GDP_c$ and $GDP_\mathrm{world}$ are the GDP of country $c$ and the world.

For both metrics, a value of 1 indicates representation proportional to the country's population (for $\psi_c$) or economic output (for $\phi_c$). Values greater than 1 indicate overrepresentation, and values below 1 indicate underrepresentation relative to the baseline. These normalizations facilitate meaningful cross-country comparisons despite substantial differences in population size and economic capacity, as shown in Figure~\ref{fig:worldmap-normalized}.

\textbf{Analysis of inferred author sex.}  \label{subsec:sexinference}
As a secondary, exploratory analysis, we extracted the first and last authors from each paper. In some cases, the last author listed was ``ADNI'',  referring to the dataset, project, or consortium. To address this, we manually corrected 39 such instances by consulting DOI links to identify and retrieve the individual author listed prior to ``ADNI''. We acknowledge that gender and sex are distinct concepts and neither is strictly binary. Although many available tools claim to estimate gender, gender can only be self-identified. In practice, these tools instead attempt to infer sex assigned at birth based on individuals' names. For this purpose, we used Wiki-Gendersort~\cite{berube2020wiki}, an open-source tool for name-based sex inference, which has shown comparable performance \cite{sebo2021performance} to proprietary services like Gender API~\cite{genderapi} or genderize.io~\cite{genderizeio}). This approach has important limitations: accuracy is not uniform across name origins, and it is likely lower for names from regions underrepresented in the training data used to build such tools, a concern directly relevant to a study of representation. We therefore report these results as indicative of a broader pattern rather than as a precise measurement, and we do not treat inferred sex as a primary variable elsewhere in this paper. Table~\ref{tab:authorship_stats} shows the number of authors per year $N_{authors}$, and the number discarded $N_{discarded}$ due to initials (INI), unisex (UNI), or unknown (UNK) names.

\begin{figure*}[t]
    \centering
    \includegraphics[width=.8\linewidth]{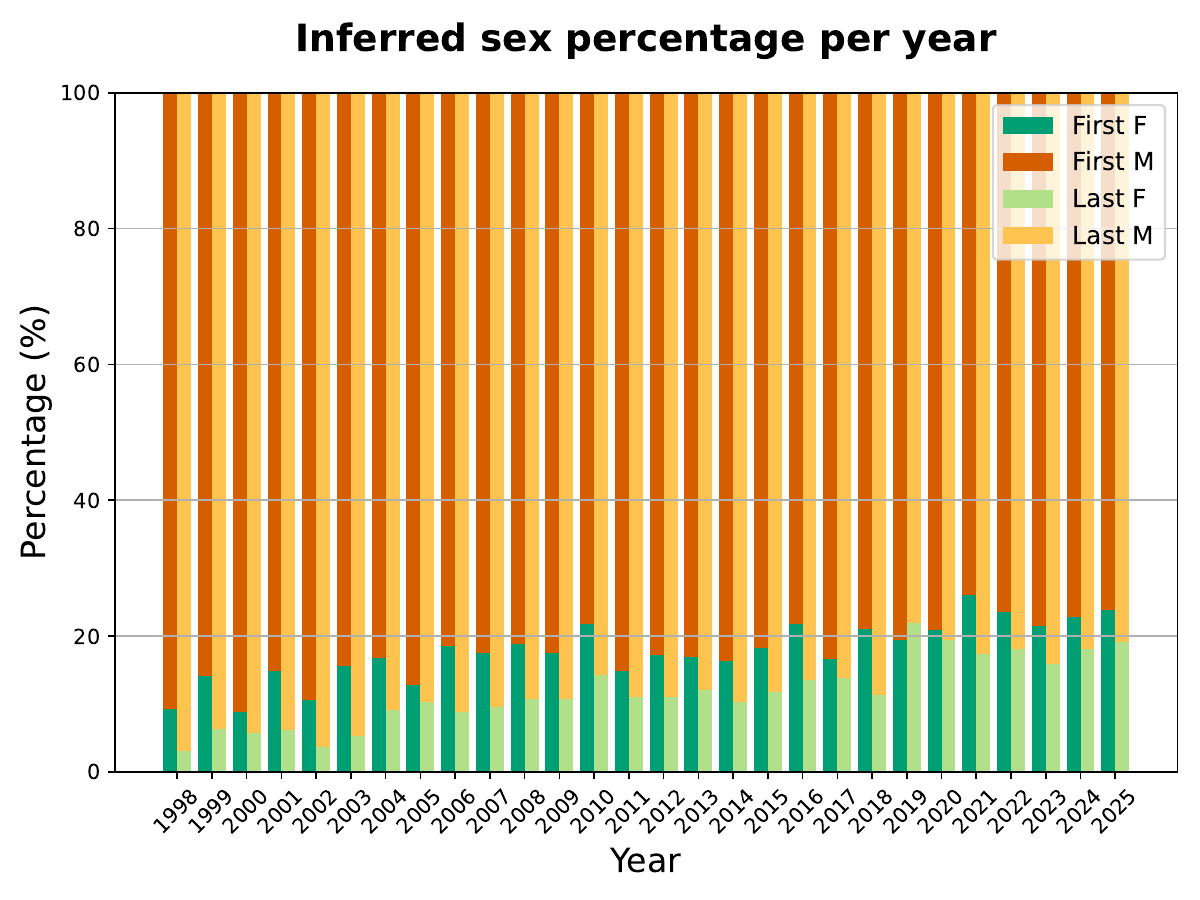}
    \caption{Sex differences represented by yearly percentages. Female authors remain underrepresented in both first- and last-authorship positions, although the disparity has gradually narrowed over time. First F: first author is female, First M: first author is male, Last F: last author is female, Last M: last author is male.}
    \label{fig:sexdiff}
\end{figure*}

Fig.~\ref{fig:sexdiff} reports the count of inferred female and male authors, $N_{inferred}$. This plot highlights a pronounced imbalance for both first and last authorship positions. Although male authors were once overwhelmingly dominant, comprising about 95\% of the total, this disparity has decreased over time, with their share dropping roughly to 80\% in more recent years. While female representation has increased in both first and last authorship positions, the proportion remains consistently lower for last authors, indicating more pronounced and persistent male overrepresentation in senior and leadership roles.

\renewcommand{\arraystretch}{1.2}  

\begin{table}[h]
\centering
\caption{Summary of authorship statistics by year. We report first/last author counts. Abbreviations: INI, initials; UNI, unisex; UNK, unknown.}
\label{tab:authorship_stats}
\begin{tabular}{lccccccccc}
\toprule
 &  & \multicolumn{4}{c}{Discarded} &  \multicolumn{3}{c}{Inferred sex}  \\
\cmidrule(lr){3-6} \cmidrule(lr){7-9}
Year & $N_{authors}$ & INI & UNI & UNK & $N_{discarded}$ & Female & Male & $N_{inferred}$ \\
\midrule
1998 & 132 & 30 / 33 & 4 / 1 & 1 / 1 & 35 / 35 & 9 / 3 & 88 / 94 & 97 / 97 \\
1999 & 133 & 53 / 53 & 1 / 0 & 1 / 0 & 55 / 53 & 11 / 5 & 67 / 74 & 78 / 79 \\
2000 & 133 & 41 / 44 & 1 / 0 & 0 / 1 & 42 / 45 & 8 / 5 & 83 / 83 & 91 / 85\\
2001 & 258 & 59 / 64 & 4 / 0 & 6 / 2 & 69 / 66 & 28 / 12 & 161  / 180 & 189 / 192 \\
2002 & 183 & 40 / 45 & 7 / 0 & 2 / 3 & 49 / 48 & 14 / 5 & 118 / 130 & 132 / 135 \\
2003 & 236 & 45 / 47 & 10 / 1 & 2 / 0 & 57 / 48 & 27 / 10 & 152 / 178 & 179 / 188 \\
2004 & 268 & 8 / 13 & 4 / 4 & 5 / 0 & 17 / 17 & 42 / 23 & 209 / 228 & 251 / 251 \\
2005 & 237 & 38 / 37 & 11 / 3 & 0 / 2 & 49 / 42 & 23 / 17 & 165 / 178 & 188 / 195 \\
2006 & 232 & 6 / 11 & 12 / 4 & 4 / 2 & 22 / 17 & 38 / 19 & 172 / 196 & 210 / 215 \\
2007 & 237 & 28 / 21 & 16 / 2 & 3 / 6 & 47 / 29 & 33 / 19 & 157 / 189 & 190 / 208 \\
2008 & 258 & 3 / 4 & 20 / 7 & 8 / 5 & 31 / 16 & 39 / 25 & 188 / 217 & 227 / 242 \\
2009 & 259 & 3 / 4 & 19 / 1 & 6 / 2 & 28 / 7 & 41 / 26 & 190 / 226 & 231 / 252 \\
2010 & 251 & 24 / 29 & 16 / 2 & 10 / 4 & 50 / 35 & 43 / 30 & 158 / 186 & 201 / 216 \\
2011 & 249 & 13 / 19 & 14 / 1 & 4 / 2 & 31 / 22 & 32 / 26 & 186 / 201 & 218 / 227 \\
2012 & 252 & 19 / 21 & 15 / 2 & 5 / 4 & 39 / 27 & 37 / 27 & 176 / 198 & 213 / 225 \\
2013 & 262 & 1 / 4 & 18 / 7 & 7 / 3 & 26 / 14 & 40 / 31 & 196 / 217 & 236 / 248 \\
2014 & 253 & 1 / 6 & 15 / 12 & 0 / 2 & 16 / 20 & 41 / 24 & 196 / 209 & 237 / 233 \\
2015 & 263 & 9 / 9 & 23 / 15 & 2 / 10 & 34 / 34 &  41 / 27 & 188 / 202 & 229 / 229 \\
2016 & 228 & 5 / 8 & 27 / 10 & 3 / 4 & 35 / 22 & 39 / 28 & 154 / 178 & 193 / 206 \\
2017 & 255 & 5 / 8 & 24 / 9 & 1 / 6 & 30 / 23 & 38 / 33 & 187 / 199 & 225 / 232 \\
2018 & 372 & 4 / 10 & 34 / 29 & 10 / 6 & 48 / 45 & 65 / 40 & 259 / 287 & 324 / 327 \\
2019 & 538 & 4 / 12 & 72 / 49 & 6 / 10 & 82 / 71 & 88 / 104 & 368 / 363 & 456 / 467 \\
2020 & 542 & 0 / 9 & 84 / 63 & 8 / 9 & 92 / 81 & 94 / 91 & 356 / 370 & 450 / 461 \\
2021 & 531 & 1 / 13 & 71 / 55 & 10 / 5 & 82 / 73 & 118 / 79 & 331 / 379 & 449 / 458 \\
2022 & 573 & 0 / 11 & 85 / 71 & 12 / 7 & 97 / 89 & 113 / 88 & 363 / 396 & 476 / 484 \\
2023 & 730 & 3 / 11 & 103 / 84 & 10 / 6 & 116 / 101 & 132 / 100 & 482 / 529 & 614 / 629 \\
2024 & 856 & 1 / 9 & 129 / 79 & 13 / 7 & 143 / 95 & 163 / 137 & 550 / 624 & 713 / 761 \\
2025 & 1027 & 1 / 4 & 158 / 126 & 11 / 8 & 170 / 138 & 203 / 171 & 654 / 718 & 857 / 889 \\
\bottomrule
\end{tabular}
\end{table}

\clearpage
\newpage
\subsection{Data Cards} \label{subsec:datacards}
We accompany the \textbf{Atlas of MICCAI (1998-2025) dataset} with a Datasheet~\cite{Gebru2021datasheets}, which documents the collected dataset, consisting of metadata from MICCAI conference papers. It outlines the dataset's composition, including metadata and content extracted from published papers, and explains the collection process from official conference proceedings. The document further summarizes preprocessing, cleaning, and labeling steps to ensure consistency and usability, and discusses potential research and analysis use cases. Finally, it provides information on data distribution, access, and ongoing maintenance to support transparency and reproducibility.

\subsubsection{Motivation}
\textbf{For what purpose was the dataset created?}
The dataset was developed to examine the composition of MICCAI over the years, focusing on affiliations, types of affiliations, geographic distribution, and research topics.

\textbf{Who created the dataset (e.g., which team, research group) and on behalf of which entity (e.g., company, institution, organization)?}
The dataset was created by Amelia \JimSan{} (AJS) at University of Barcelona.

\textbf{Who funded the creation of the dataset?}
AJS has received funding under the Horizon Europe grant 101213369 (DVPS).


\begin{table}[ht]
\centering
\caption{Data description.}
\label{tab:datadescript}
\small
\begin{tabular}{L{0.02\columnwidth}|L{0.38\columnwidth}|L{0.6\columnwidth}
}
& Field & Description \\ \toprule
\multirow{16}{*}{\rotatebox{90}{\textbf{Raw data}}} 
& \texttt{conf\_name} & Name of the conference \\
& \texttt{conf\_location} & Country where the conference was held \\
& \texttt{year} & Year in which the conference was held \\
& \texttt{n\_volumes} & Number of volumes in the proceedings \\
& \texttt{n\_papers} & Number of papers in the proceedings \\
& \texttt{chapter\_url} & Link to the conference chapter \\
& \texttt{doi\_url} & Digital Object Identifier (DOI) of the paper \\
& \texttt{title} & Title of the paper \\
& \texttt{abstract} & Abstract of the paper \\
& \texttt{keywords} & Keywords for the paper \\
& \texttt{metrics} & Bibliometric data dictionary that contains: \\
& \quad |-- \texttt{accesses} & |-- Number of accesses \\ 
& \quad |-- \texttt{citations} & |-- Number of citations \\ 
& \quad |-- \texttt{almetrics} & |-- Altmetrics \\ 
& \texttt{ack} & Acknowledgments \\
& \texttt{authors} & List of authors \\
& \texttt{authors\_with\_numbers} & List of authors with numeric tags representing their affiliation IDs.\\
& \texttt{affiliations} & Dictionary where each affiliation is a key, and each key maps to the affiliation's raw \texttt{address} along with the list of \texttt{author}s linked to it \\ 
\midrule
\multirow{18}{*}{\rotatebox{90}{\textbf{Annotated data}}} 
& \texttt{affiliations\_annotated} & Dictionary of cleaned affiliations, with one entry per affiliation that contains: \\
& \quad |-- \texttt{institution} & |-- Name of the institution \\ 
& \quad |-- \texttt{country} & |-- ISO 31166-1 alpha-3 format code of the affiliated country \\ 
& \quad |--\texttt{institution\_category} & |-- Classification category based on the ROR taxonomy (\textit{education, healthcare, facility, company} \\
& \quad |-- \texttt{economic\_category} & |-- Economic category according to the UN classification \\
& \quad |-- \texttt{author} & |-- List of authors. 
\\ 
& \texttt{first\_author} & Name of the first author \\
& \texttt{first\_country} & Country where the first author is affiliated \\
& \texttt{first\_region} & Location 1 based on the UN's geographical categorization (e.g., Africa) \\
& \texttt{first\_subregion} & Location 2 based on the UN's geographical categorization (e.g., North Africa) \\
& \texttt{first\_aff\_cat} & ROR category for the institution where the first author is affiliated \\
& \texttt{first\_aff\_eco} & UN economic category for the country where the first author is affiliated \\
& \texttt{first\_aff\_eco\_gni} & UN GNI-based economic category for the country where the first author is affiliated \\
& \texttt{list\_regions} & List of UN Location 1 regions in a paper \\
& \texttt{list\_subregions} & List of UN Location 2 regions in a paper  \\
& \texttt{list\_affiliations} & List of ROR affiliation types in a paper \\
& \texttt{list\_countries} & List of countries in a paper \\
& \texttt{list\_economic\_categories} & List of UN economic categories in a paper \\
& \texttt{list\_economic\_categories\_gni} & List of GNI-based economic categories in a paper \\
\midrule
\multirow{2}{*}{\rotatebox{90}{\textbf{Research topics}}} 
& \texttt{categories} & List of selected research categories for papers between 2021 and 2025, including \textit{Applications}, \textit{Clinical Applications}, \textit{Machine Learning}, \textit{Modalities}, \textit{Surgery \& Intervention}. \\ 
& \quad |-- \texttt{subcategories} & List of selected research subcategories for papers between 2021 and 2025. Full list in Appendix~\ref{subsec:categories_2021_2025}\\ 
\end{tabular}
\end{table}

\subsubsection{Composition}

\textbf{What do instances that comprise the dataset represent (e.g., documents, photos, people, countries)?} Instances represent papers at MICCAI conferences, there is information about the authors, affiliation and countries. 

\textbf{How many instances are there in total (of each type, if appropriate)?} 9,748. In the affiliation analysis, we discarded four papers that did not contain affiliation information.

\textbf{Does the dataset contain all possible instances or is it a sample (not necessarily random) of instances from a larger set?} We retrieved all possible instances from the main conference. 

\textbf{What data does each instance consist of?} As shown in Table~\ref{tab:datadescript}, each instance consists of conference name, conference location, year, number of volumes in the proceedings, number of papers in the proceedings, link to the conference chapter, paper DOI, title of the paper, abstract of the paper, keywords for the paper, bibliometric data dictionary (containing the number of accesses, citations, and altmetrics), acknowledgments, the list of authors, the numbered list of authors (numbers correspond to their affiliation IDs), and an affiliation dictionary with a key for each affiliation. For papers published between 2021 and 2025, instances also include information on research categories.

\textbf{Is there a label or target associated with each instance?} The field \texttt{affiliations\_annotated} contains the affiliation labels associated with each paper. For papers published between 2021 and 2025, the fields \texttt{categories} and \texttt{subcategories} contain information on research categories. Additional labels provide geographical and economic classification according to the UN framework. Details of all the labels are provided in Table~\ref{tab:datadescript}. 

\textbf{Is any information missing from individual instances?} Affiliation information is missing for 4 papers, which were discarded for the analysis.

\textbf{Are there recommended data splits (e.g., training, development/validation,
testing)?} No. No experiments were conducted, and no models were trained.

\textbf{Is the dataset self-contained, or does it link to or otherwise rely on
external resources (e.g., websites, tweets, other datasets)?} The dataset is self-contained. 

\textbf{Does the dataset contain data that might be considered confidential (e.g., data that is protected by legal privilege or by doctor–
patient confidentiality, data that includes the content of individuals’ non-public communications)? } No, all data collected was publicly available at Springer Nature Link website.

\textbf{Does the dataset contain data that, if viewed directly, might be offensive, insulting, threatening, or might otherwise cause anxiety?} No.

\subsubsection{Collection Process}

\textbf{How was the data associated with each instance acquired?} We gathered all available metadata from the Springer Nature Link website using the Python package Beautiful Soup~\cite{richardson2007beautiful}. Details provided in Section~\ref{subsec:datacollection}.

\textbf{Who was involved in the data collection process (e.g., students, crowdworkers, contractors) and how were they compensated (e.g., how much were crowdworkers paid)?} One of the authors carried out the data collection process; no compensation was involved. 

\textbf{Over what timeframe was the data collected?} Data was collected between November of 2025 and September of 2026. The data correspond to MICCAI papers from 1998 to 2025.

\textbf{Were any ethical review processes conducted (e.g., by an institutional review board)?} No. 

\subsubsection{Preprocessing/cleaning/labeling}

\textbf{Was any preprocessing/cleaning/labeling of the data done (e.g., discretization or bucketing, tokenization, part-of-speech tagging, SIFT feature extraction, removal of instances, processing of missing values)? } Affiliations were annotated as described in Section~\ref{subsec:datacollection}.

\textbf{Was the ``raw'' data saved in addition to the preprocessed/cleaned/labeled data (e.g., to support unanticipated future uses)?} Yes, ``raw'' \texttt{affiliations} were also saved in the json file.

\textbf{Is the software that was used to preprocess/clean/label the data available?} Yes, the code is available at this GitHub repository \url{https://github.com/ameliajimenez/atlas-of-miccai}.

\subsubsection{Uses}

\textbf{Has the dataset been used for any tasks already?} The dataset was used to map the institutional and geographic trends of MICCAI at this paper.

\textbf{Is there a repository that links to any or all papers or systems that
use the dataset?} Yes, this GitHub repository \url{https://github.com/ameliajimenez/atlas-of-miccai}.

\textbf{What (other) tasks could the dataset be used for?} The dataset could be used to investigate additional dimensions of representations, influence, and power. Incorporating bibliometric and other network analysis techniques, such as citation patterns, could help uncover how influence and knowledge across regions and institutions. Furthermore, analyzing acknowledgments and funding sources would provide insights into the structural factors shaping research production, collaboration, and visibility, enabling a more comprehensive understanding of equity and participation in the field. 

\textbf{Are there tasks for which the dataset should not be used?} The dataset should not be used for tasks requiring sensitive, private or non-public information, as it contains only publicly available conference papers and their metadata. It is not suitable for applications that attempt to extract confidential information beyond what is explicitly published in the papers.

\subsubsection{Distribution}

\textbf{How will the dataset will be distributed (e.g., tarball on website, API, GitHub)?} The dataset will be available on Zenodo. 

\textbf{Will the dataset be distributed under a copyright or other intellectual property (IP) license, and/or under applicable terms of use (ToU)?} The dataset is under CC-BY-NC license.

\subsubsection{Maintenance}

\textbf{Who will be supporting/hosting/maintaining the dataset?} Amelia \JimSan{}.

\textbf{How can the owner/curator/manager of the dataset be contacted (e.g., email address)?} amelia.jimenez @ ub.edu.

\textbf{Will the dataset be updated (e.g., to correct labeling errors, add new instances, delete instances)?} If errors are reported, the dataset will be updated on Zenodo, which supports version control.

\newpage
\clearpage
\subsection{Prompt for Structured Affiliation Extraction} \label{subsec:prompt}
Our prompt is adapted from \url{https://github.com/juansgomez87/ismir-papers} \cite{gomez2025beyond}. 

We designed the prompt to extract structured affiliation information from unstructured text, such as author affiliation strings in conference papers. It guides the model to identify and normalize key entities, including institution names, and countries, while handling variations in formatting. We further refined the prompt to improve consistency and robustness across diverse affiliation styles observed in MICCAI papers. This adaptation ensures reliable parsing and facilitates downstream analysis of institutional and geographic trends.

\begin{lstlisting}[language=Python]
    
def get_clean_affiliations_with_llm(affiliations_list):
    client = InferenceClient(
        provider="groq",
        api_key="",
    )

    completion = client.chat.completions.create(
        model="meta-llama/Llama-4-Scout-17B-16E-Instruct",
        messages = [
            {
                "role": "user",
                "content": "You will be given a list of affiliation names as a string separated by ';'. Extract:"
                           "1. Affiliation name (institution only, not departments)"
                           "2. Affiliation country"
                           "3. Affiliation category"
                           
                            "The categories are strictly:"
                           "- education (e.g., university)"
                           "- company (e.g., commercial entities)"
                           "- facility (e.g., research center, governmental agencies, non-profit, hospital)"
                                                    
                           "Important rules" 
                           "1. Ignore departments, laboratories, and groups. An affiliation should only be something such as a university, company, or similar entity"
                           "2. Countries must be in ISO 31166-1 A3 format, which are always 3 letters (Sweden should be SWE, not SE)"
                           "3. If country cannot be determined, use 'Unknown' for the country ISO code"
                           "4. If the category cannot be determined, use 'Unknown' for the category"
                           "5. Ommit articles for affiliations (e.g. use University of Tokyo, not the University of Tokyo)"
                           
                           "The text may have formatting issues that you need to handle: "
                           "1. Joined words (e.g., 'JohnSmith' $\rightarrow$ 'John Smith', 'StanfordUniversity' $\rightarrow$ 'Stanford University')"
                           "2. Special symbols for affiliations (e.g., $\flat$, $\natural$, $\dagger$, *, [1], 1)"
                           "3. Abbreviations (e.g. 'Univ' $\rightarrow$  'University', 'Inst.' $\rightarrow$  'Institute'"
                        
                           "Examples:"
                           "Input: "
                           "Surgical Planning Laboratory, Brigham and Women's Hospital, Harvard Medical School, 75, Francis Street, 02115, Boston, MA, USA"
                           "MIT AI Laboratory, Cambridge, MA, USA,;"
                           "Siemens Corporate Research, Inc., 755 College Road East, 08540, Princeton, NJ,;"
                           "Image Sciences Institute, room E 01.334, University Hospital Utrecht, Heidelberglaan 100, 3584, CX Utrecht, the Netherlands;"
                           "Lab. of Experimental Surgery, Hadassah Univ. Hospital, 91120, Jerusalem, Israel"
                           
                           "Output: 
                           Harvard Medical School>USA>education;
                           MIT>USA>education; 
                           Siemens>USA>company; 
                           University Hospital Utrecht>NLD>healthcare;
                           Hadassah Univ. Hospital>ISR>healthcare; "    
                           
                           "Make sure the values are always strings, and verify your response matches these technical specifications exactly. "
                           "Respond only with a list of strings separated by ';', with the information without a header nor explanation."

                           "Give exactly one affiliation for each entry, each entry is separated by ';':" + ';'.join(affiliations_list)
            }
        ],
    )

    affiliations_string = completion.choices[0].message.content.replace('\n','')
    affiliations_string = [aff.strip() for aff in affiliations_string.split(';')]
    affiliations_list = list(filter(None, affiliations_string))
    return affiliations_list
\end{lstlisting}

\newpage
\clearpage
\subsection{List of categories and subcategories from 2021 to 2025 by year} \label{subsec:categories_2021_2025}
\section*{MICCAI 2021}
\noindent Source: \url{https://miccai2021.org/openaccess/paperlinks/categories/index.html} (531 papers)
\begin{itemize}
  \item \textbf{Applications}
  \begin{itemize}
    \item Computational (Integrative) Pathology
    \item Computational Anatomy and Physiology
    \item Computer Aided Diagnosis
    \item Image Reconstruction
    \item Image Registration
    \item Image Segmentation
    \item Integration of Imaging with Non-Imaging Biomarkers
    \item Outcome/disease prediction
    \item Population Imaging and Imaging Genetics
    \item Visualisation in Biomedical Imaging
  \end{itemize}
  \item \textbf{Clinical Applications}
  \begin{itemize}
    \item Abdomen
    \item Breast
    \item Cardiac
    \item Dermatology
    \item Fetal Imaging
    \item Lung
    \item Musculoskeletal
    \item Neuroimaging - Brain Development
    \item Neuroimaging - DWI and Tractography
    \item Neuroimaging - Functional Brain Networks
    \item Neuroimaging - Others
    \item Oncology
    \item Ophthalmology
    \item Vascular
  \end{itemize}
  \item \textbf{Machine Learning}
  \begin{itemize}
    \item Active Learning
    \item Attention models
    \item Domain adaptation
    \item Interpretability / Explainability
    \item Reinforcement learning
    \item Self-supervised learning
    \item Semi-supervised learning
    \item Uncertainty
    \item Weakly supervised learning
  \end{itemize}
  \item \textbf{Modalities}
  \begin{itemize}
    \item Biophotonics
    \item CT
    \item EEG/ECG
    \item Histopathology
    \item MRI
    \item Microscopy
    \item PET/SPECT
    \item Text (clinical/radiology reports)
    \item Ultrasound
    \item Video
    \item other
  \end{itemize}
  \item \textbf{Surgery \& Intervention}
  \begin{itemize}
    \item Image-Guided Interventions and Surgery
    \item Interventional Imaging Systems
    \item Interventional Simulation Systems
    \item Medical Robotics and Haptics
    \item Surgical Data Science
    \item Surgical Planning and Simulation
    \item Surgical Skill and Work Flow Analysis
    \item Surgical Visualization and Mixed, Augmented and Virtual Reality
  \end{itemize}
\end{itemize}

\section*{MICCAI 2022}
\noindent Source: \url{https://conferences.miccai.org/2022/papers/categories/} (572 papers)
\begin{itemize}
  \item \textbf{Applications}
  \begin{itemize}
    \item Computational (Integrative) Pathology
    \item Computational Anatomy and Physiology
    \item Computer Aided Diagnosis
    \item Image Reconstruction
    \item Image Registration
    \item Image Segmentation
    \item Integration of Imaging with Non-Imaging Biomarkers
    \item Outcome/disease prediction
    \item Population Imaging and Imaging Genetics
    \item Visualisation in Biomedical Imaging
  \end{itemize}
  \item \textbf{Organ}
  \begin{itemize}
    \item Abdomen
    \item Brain
    \item Breast
    \item Eye
    \item Fetal
    \item Heart
    \item Lung
    \item Musculoskeleton
    \item Other
    \item Skin
    \item Tumor
    \item Vessel
  \end{itemize}
  \item \textbf{Machine Learning}
  \begin{itemize}
    \item Active Learning
    \item Continual Learning
    \item Data efficient Learning
    \item Interpretability / Explainability
    \item Meta-learning
    \item Model Generalizability
    \item Other
    \item Reinforcement learning
    \item Semi-/Weakly-/Un-/Self-supervised representation learning
    \item Transfer learning
    \item Uncertainty
  \end{itemize}
  \item \textbf{Modalities}
  \begin{itemize}
    \item Biophotonics
    \item CT
    \item EEG/ECG
    \item Endoscope
    \item Histopathology
    \item MRI
    \item Microscopy
    \item PET/SPECT
    \item Text (clinical/radiology reports)
    \item Ultrasound
    \item Video
    \item other
  \end{itemize}
  \item \textbf{Surgery \& Intervention}
  \begin{itemize}
    \item Image Analysis in Robot-Assisted Surgery
    \item Image-Guided Interventions
    \item Interventional Imaging Systems
    \item Interventional Simulation Systems
    \item Medical Robotics and Haptics
    \item Surgical Data Science
    \item Surgical Planning and Simulation
    \item Surgical Scene Understanding
    \item Surgical Skill and Work Flow Analysis
    \item Surgical Visualization and Mixed/Augmented/Virtual Reality
  \end{itemize}
\end{itemize}

\section*{MICCAI 2023}
\noindent Source: \url{https://conferences.miccai.org/2023/papers/categories/} (730 papers)
\begin{itemize}
  \item \textbf{Applications}
  \begin{itemize}
    \item Computational (Integrative) Pathology
    \item Computational Anatomy and Physiology
    \item Computer Aided Diagnosis
    \item Image Reconstruction
    \item Image Registration
    \item Image Segmentation
    \item Integration of Imaging with Non-Imaging Biomarkers
    \item Population Imaging and Imaging Genetics
    \item Rigorous Evaluations of Methodology in Clinical Workflows
    \item Treatment Response and Outcome/Disease Prediction
    \item Visualization in Biomedical Imaging
  \end{itemize}
  \item \textbf{Clinical Applications}
  \begin{itemize}
    \item Abdomen
    \item Breast
    \item Cardiac
    \item Dermatology
    \item Fetal Imaging
    \item Lung
    \item Musculoskeletal
    \item Neuroimaging - Brain Development
    \item Neuroimaging - DWI and Tractography
    \item Neuroimaging - Functional Brain Networks
    \item Neuroimaging - Others
    \item Oncology
    \item Ophthalmology
    \item Vascular
  \end{itemize}
  \item \textbf{Machine Learning}
  \begin{itemize}
    \item Active Learning
    \item Attention models
    \item Continual Learning
    \item Data Efficient Learning
    \item Interpretability / Explainability
    \item Meta-learning
    \item Model Generalizability / Federated Learning
    \item Other
    \item Reinforcement Learning
    \item Semi-/Weakly-/Un-/Self-supervised Representation Learning
    \item Transfer learning
    \item Uncertainty
  \end{itemize}
  \item \textbf{Modalities}
  \begin{itemize}
    \item Biophotonics
    \item CT
    \item EEG/ECG
    \item Histopathology
    \item MRI
    \item Microscopy
    \item PET/SPECT
    \item Text (clinical/radiology reports)
    \item Ultrasound
    \item Video
    \item other
  \end{itemize}
  \item \textbf{Surgery \& Intervention}
  \begin{itemize}
    \item Image-Guided Interventions and Surgery
    \item Interventional Imaging Systems
    \item Interventional Simulation Systems
    \item Medical Robotics and Haptics
    \item Surgical Data Science
    \item Surgical Planning and Simulation
    \item Surgical Scene Understanding
    \item Surgical Skill and Work Flow Analysis
    \item Surgical Visualization and Mixed/Augmented/Virtual Reality
  \end{itemize}
\end{itemize}

\section*{MICCAI 2024}
\noindent Source: \url{https://papers.miccai.org/miccai-2024/categories/} (856 papers)
\begin{itemize}
  \item \textbf{Applications}
  \begin{itemize}
    \item Computational (Integrative) Pathology
    \item Computational Anatomy and Physiology
    \item Computer Aided Diagnosis, Treatment Response, and Outcome Prediction
    \item Image Formation and Reconstruction
    \item Image Registration
    \item Image Segmentation
    \item Integration of Imaging with Non-Imaging Biomarkers
    \item Population Imaging and Imaging Genetics
    \item Visualization in Biomedical Imaging
  \end{itemize}
  \item \textbf{Clinical Applications}
  \begin{itemize}
    \item Abdomen
    \item Breast
    \item Cardiac
    \item Dermatology
    \item Fetal Imaging
    \item Lung
    \item Musculoskeletal
    \item Neuroimaging - Brain Development
    \item Neuroimaging - DWI and Tractography
    \item Neuroimaging - Functional Brain Networks
    \item Neuroimaging - Others
    \item Oncology
    \item Ophthalmology
    \item Vascular
  \end{itemize}
  \item \textbf{Health Equity \& Clinical Translation}
  \begin{itemize}
    \item Biomedical Image Computing for Neglected Diseases
    \item Clinical Implementation and Monitoring of Imaging Solutions
    \item Human-centred AI in Medical Imaging
    \item Image-based Personalised Medicine
    \item Imaging Solutions for Vulnerable and Under-represented Populations
    \item Imaging-related Clinical Studies
    \item Low-cost and Point-of-care Imaging Solutions
    \item MIC and CAI Solutions for Minimally Trained Healthcare Workers
    \item MIC and CAI Solutions for Remote and Rural Regions
    \item MIC and CAI for Limited-resource Settings
    \item Telemedicine and Mobile Health Imaging Technologies
  \end{itemize}
  \item \textbf{Machine Learning}
  \begin{itemize}
    \item Active Learning
    \item Algorithmic Fairness
    \item Attention models
    \item Continual Learning
    \item Data Efficient Learning
    \item Foundation Models
    \item Interpretability / Explainability
    \item Meta-learning
    \item Model Generalizability / Federated Learning
    \item Other
    \item Reinforcement Learning
    \item Semi-/Weakly-/Un-/Self-supervised Representation Learning
    \item Transfer Learning
    \item Uncertainty
  \end{itemize}
  \item \textbf{Modalities}
  \begin{itemize}
    \item Biophotonics
    \item CT
    \item EEG/ECG
    \item Endoscope
    \item Histopathology
    \item Integration of Imaging and Non-imaging Data
    \item MRI
    \item Microscopy
    \item PET/SPECT
    \item Ultrasound
    \item Video
    \item other
  \end{itemize}
  \item \textbf{Surgery \& Intervention}
  \begin{itemize}
    \item Image-Guided Interventions and Surgery
    \item Interventional Imaging Systems
    \item Interventional Simulation Systems
    \item Medical Robotics and Haptics
    \item Surgical Data Science
    \item Surgical Planning and Simulation
    \item Surgical Scene Understanding
    \item Surgical Skill and Work Flow Analysis
    \item Surgical Visualization and Mixed/Augmented/Virtual Reality
  \end{itemize}
\end{itemize}

\section*{MICCAI 2025}
\noindent Source: \url{https://papers.miccai.org/miccai-2025/categories/} (1027 papers)
\begin{itemize}
  \item \textbf{Applications}
  \begin{itemize}
    \item Anomaly Detection
    \item Brain Network Analysis
    \item Computational (Integrative) Pathology
    \item Computational Anatomy and Physiology
    \item Computer Aided Diagnosis
    \item Image Registration
    \item Image Segmentation
    \item Image Synthesis / Augmentation / Super-Resolution
    \item Image-Guided Interventions
    \item Integration of Imaging with Non-Imaging Biomarkers
    \item Other
    \item Visualization in Biomedical Imaging
  \end{itemize}
  \item \textbf{Body}
  \begin{itemize}
    \item Abdomen
    \item Brain
    \item Breast
    \item Cardiac
    \item Eye
    \item Fetal / Pediatric Imaging
    \item Gynecology
    \item Lung
    \item Skin
    \item Spine
    \item Urology
    \item Vasculature
    \item other
  \end{itemize}
  \item \textbf{Machine Learning}
  \begin{itemize}
    \item Algorithmic Fairness
    \item Deep Learning
    \item Domain Adaptation / Harmonization
    \item Foundation Models
    \item Interpretability / Explainability
    \item Other
    \item Semi- / Weakly- / Self-supervised Learning
    \item Uncertainty
    \item Validation
  \end{itemize}
  \item \textbf{Modalities}
  \begin{itemize}
    \item CT / X-Ray
    \item EEG / ECG
    \item Endoscopy
    \item MRI
    \item MRI - Diffusion Imaging
    \item MRI - Functional MRI
    \item Microscopy
    \item Nuclear Imaging
    \item Other
    \item Photograph / Video
    \item Robotics
    \item Spectroscopy
    \item Ultrasound
  \end{itemize}
  \item \textbf{Special Topic}
  \begin{itemize}
    \item Biomedical Image Computing for Neglected Diseases
    \item Low-Cost and Point-of-Care Imaging Solutions
    \item MIC and CAI Solutions for Limited-Resource Environments
    \item Telemedicine and Mobile Health Imaging Technologies
  \end{itemize}
  \item \textbf{Surgery}
  \begin{itemize}
    \item Data Science
    \item Mixed / Augmented / Virtual Reality
    \item Navigation
    \item Other
    \item Planning and Simulation
    \item Scene Understanding
    \item Skill and Work Flow Analysis
  \end{itemize}
\end{itemize}


\end{document}